\documentclass{emulateapj-rtx4}
\usepackage{subfigure}
\usepackage{floatflt,graphicx}
\usepackage{lscape}

\RequirePackage{epsf,graphicx}

\newcommand{\HI}{H\,{\sc i}}
\newcommand{\HII}{H\,{\sc ii}}

\newcommand{\arcs}{\arcsec }

\slugcomment{Accepted for Publication in the Astrophysical Journal}

\shorttitle{\HI-LIRG HIZOA J0836-43}
\shortauthors{Cluver et al.}

\begin{document}

\title{Active Disk Building in a local \HI-Massive LIRG: The Synergy between Gas, Dust, and Star Formation}

\author{M.E. Cluver\altaffilmark{1,2}, T.H. Jarrett\altaffilmark{1}, R.C. Kraan-Korteweg\altaffilmark{2}, B.S. Koribalski\altaffilmark{3}, P.N. Appleton\altaffilmark{4}, J. Melbourne\altaffilmark{5}, B. Emonts\altaffilmark{3}, P.A. Woudt\altaffilmark{2}}

\altaffiltext{1}{IPAC, California Institute of Technology, Pasadena, CA 91125}
\altaffiltext{2}{Astronomy Department; Centre for Astrophysics, Cosmology and Gravity, University of Cape Town, Rondebosch, 7700, South Africa}
\altaffiltext{3}{Australia Telescope National Facility, CSIRO, Epping, NSW 1710, Australia}
\altaffiltext{4}{NASA Herschel S Center, California Institute of Technology, Pasadena, CA 91125}
\altaffiltext{5}{Division of Physics, Mathematics and Astronomy, California Institute of Technology, Pasadena, CA 91125, USA}

\begin{abstract}

HIZOA~J0836-43 is the most \HI-massive ($M_{\rm{HI}} = 7.5\times10^{10}\, M_{\sun}$) galaxy detected in the HIPASS volume ($\delta = -90\arcdeg\ {\rm to} +25\arcdeg, v < 12\,700$ km\,s$^{-1}$) and lies optically hidden behind the Milky Way. Markedly different from other extreme \HI\ disks in the local universe, it is a luminous infrared galaxy (LIRG) with an actively star forming disk ($>$50\,kpc), central to its $\sim 130$\,kpc gas disk, with a total star formation rate (SFR) of $\sim 20.5\, M_{\odot}\, {\rm yr^{-1}}$. {\it Spitzer} spectroscopy reveals an unusual combination of powerful polycyclic aromatic hydrocarbon (PAH) emission coupled to a relatively weak warm dust continuum, suggesting photodissociation region (PDR)-dominated emission. Compared to a typical LIRG with similar total infrared luminosity ($L_{\rm TIR}$=10$^{11}L_{\sun}$), the PAHs in HIZOA~J0836-43 are more than twice as strong, whereas the warm dust continuum ($\lambda > 20$\micron) is best fit by a star forming galaxy with $L_{\rm TIR}$=10$^{10}L_{\sun}$. Mopra CO observations suggest an extended molecular gas component (H$_{2}$ + He $>$ 3.7$\times 10^{9} M_\sun$) and a lower limit of $\sim$64\% for the gas mass fraction; this is above average compared to local disk systems, but similar to that of $z\sim1.5$ $BzK$ galaxies ($\sim$57\%). However, the star formation efficiency (SFE = $L_{\rm IR}/L'_{\rm CO}$) for HIZOA~J0836-43 of 140 $L_\sun$ (K\,km\,s$^{-1}$\,pc$^2$)$^{-1}$ is similar to that of local spirals and other disk galaxies at high redshift, in strong contrast to the increased SFE seen in merging and strongly interacting systems. HIZOA~J0836-43 is actively forming stars and building a massive stellar disk. Its evolutionary phase of star formation ($M_{\rm stellar}$, SFR, gas fraction) compared to more distant systems suggests that it would be considered typical at redshift $z\sim1$. This galaxy provides a rare opportunity in the nearby universe for studying (at $z\sim0.036$) how disks were building and galaxies evolving at $z\sim1$, when similarly large gas fractions were likely more common.

\end{abstract}


\section{Introduction}

The mechanism by which galaxies acquire their gas and form stars is an essential component of theoretical models of galaxy formation and evolution \citep{Dek09}. However, large-scale cosmological simulations currently lack the dynamic range to resolve the internal dynamics of galaxies essential for modeling gas inflows associated with elevated star formation \citep{Rob09}. Since the physics responsible for converting gas into stars is complex, a major aspect of understanding this evolution is the relation between gas content and star formation. 

The cold \HI\ gas in galaxies represents the reservoir of material that can fuel star formation. Therefore \HI-massive disk galaxies provide an ideal laboratory to test models of disk formation, disk instability and star formation. 
However, giant gas disks ($M_{\rm{HI}} > 10^{10}\, M_{\sun}$) are relatively rare in the local ($z<0.1$) universe and the \HI\ mass function drops off steeply, e.g. at a mass of $M_{\rm{HI}} = 6 \times 10^{10}\, M_{\sun}$, the volume density is only $1.6 \times 10^{-5}$\, Mpc$^{-3}$ \citep{Zw05}. Ideally we would like to study galaxies that are gas rich and at an early stage of stellar building, as determined by their gas fraction, stellar mass and star formation rate (SFR), in order to probe a relatively ``pristine'' stage of evolution and limit the complexity of subsequent feedback systems.

The rare examples of local massive, gas-rich disks ($M_{\rm{HI}} > 5\times10^{10}\, M_{\sun}$) include galaxies like UGC 4288 and Malin 1 which have prominent bulges, but appear underevolved, harbouring extended low surface brightness disks \citep{Mcg94}. Their low SFRs \citep{Van00} suggest that they are in a quiescent phase, for example -- Malin 1 has a SFR of only $0.38\, M_{\sun}\, \rm{yr^{-1}}$ \citep{Rah07} -- providing few clues as to its past and future development.

Other examples of local \HI-rich disks show a range of properties. The barred Sc galaxy NGC 6744 has a \HI\ disk diameter of 54 kpc containing $M_{\rm{HI}} = 2.2\times10^{10}\, M_{\sun}$ \citep{Kor04}, but a SFR of only $0.11\, M_{\sun}\, \rm{yr^{-1}}$ \citep{Boh04}. Circinus is a Seyfert~2 galaxy with $ 8.1\times10^{9}\, M_{\sun}$ of \HI\  contained in a $\sim 120$ kpc disk \citep{Jo99} and experiencing a modest starburst with a SFR of $2.52\, M_{\sun}\, \rm{yr^{-1}}$ \citep{El98}. M83 is similarly \HI-rich with $ 7.7\times10^{9}\, M_{\sun}$ in a barred, $\sim 130$ kpc disk \citep{Huch81}. This late-type galaxy is experiencing significantly more star formation at a rate of $6\, M_{\sun}\, \rm{yr^{-1}}$ \citep{Walk93}. 

In strong contrast to the extreme \HI-disks, these galaxies have formed large stellar disks and, in the case of M83, a large-scale bar, and appear to be in an advanced stage of evolution.  The most massive source in the HIPASS BGC (\HI\ Parkes All-Sky Survey Brightest Galaxy Catalog), ESO390-G004, contains $M_{\rm{HI}} = 3.7\times10^{10}\, M_{\sun}$ and shows clear signs of interaction, but has a very low SFR, $< 0.14\, M_{\sun}\, \rm{yr^{-1}}$ \citep{Jut10}. It seems apparent that converting \HI\ fuel into stars can be challenging and the process by which galaxies accrete and retain their gas, and subsequently form stars, remains poorly understood.

Large surveys probing \HI\ content, stellar mass and SFR in the $z<0.05$ universe (e.g. HIPASS, Koribalski et al. 2004, GASS; Galex Arecibo SDSS Survey, Catinella et al. 2010) provide insight as to how cold gas responds to physical conditions in galaxies, and probes mechanisms responsible for regulating gas accretion and suppressing further galaxy growth from gas converting into stars.

\begin{figure*}[!t]
\begin{center}
\includegraphics[width=14cm]{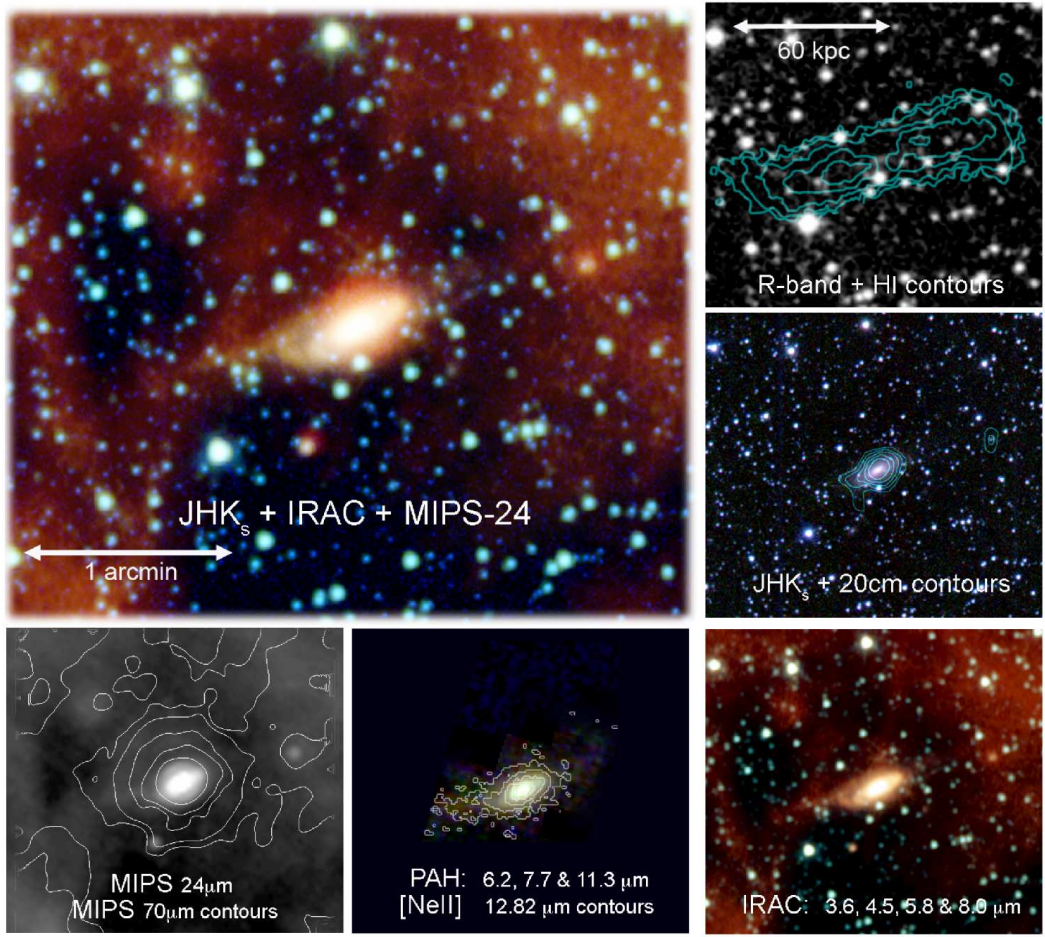}
\caption{Composite image showing HIZOA~J0836-43 at optical ($DSS$ R-band) and infrared wavelengths; in addition \HI\ (FWHM $\sim20$\arcs), 20-cm continuum (FWHM $\sim8$\arcs), $[$Ne\,{\sc ii}$]$\,12.81\micron\ (FWHM $\sim4$\arcs) and 70\micron\ (FWHM $\sim18$\arcs) contours are shown. All images have a field of view of $\sim$3\arcmin (or $\sim$130 kpc). \HI\ and 20-cm radio continuum contours are from \citet{Don06}. $JHK_s$ imaging (FWHM $\sim1.2$\arcs) is from Cluver et al. (2008).}
\label{fig:im}
\end{center}
\end{figure*}

However, studies of the early stages of stellar disk building, expected to be ubiquitous at higher redshift, are severely hampered due to observational sensitivity limits. There is growing evidence for the existence of a significant population of systems at $z>1$ with star-forming disks, where gas consumption, and not major mergers, is driving stellar mass growth (Daddi et al. 2008, Genzel et al. 2010). Ideally we should look for counterparts in the local universe where observations are easier. The recent discovery of a nearby \HI-massive galaxy undergoing a powerful starburst as it builds its stellar disk \cite[][hereafter Paper I]{Clu08}, provides a rare opportunity to study this fundamental process.

HIZOA J0836--43 is the most \HI-massive galaxy ($M_{\rm{HI}} = 7.5\times10^{10}\, M_{\sun}$) in HIPASS, discovered as part of a deep survey of the Zone of Avoidance (Kraan-Korteweg et al. 2005). Lying at a redshift of $z=0.036$, the \HI\ disk is $\sim 130$\,kpc in diameter, comparable to Malin 1, and has a dynamical mass of 1.4$\times10^{12}\, M_{\odot}$ \citep{Don06}.
However, instead of finding a relatively quiescent or even low surface brightness galaxy, typical of giant \HI\ disks in the local universe, {\it Spitzer} Space Telescope observations revealed a luminous starburst ($L_{\rm{TIR}} = 1.2 \times 10^{11} L_{\sun}$), with spatially extended star formation. The star formation rate (SFR) derived from the far-infrared (FIR) luminosities is $\sim 20.5\, M_{\sun}\, \rm{yr^{-1}}$ (Paper I), but the 20-cm radio continuum derived rate is $\sim 35\, M_{\sun}\, \rm{yr^{-1}}$ \citep{Don06}. Intriguingly, the galaxy lacks a prominent warm dust component continuum seen in other locally observed starburst systems (Paper I).

Figure 1 is a composite view of HIZOA~J0836-43 showing infrared, PAH (polycyclic aromatic hydrocarbon) and optical emission ($DSS$ R-band), as well as \HI, 20-cm continuum, $[$Ne\,{\sc ii}$]$\,12.81\micron\ and 70\micron\ contours (detailed photometric properties are presented in Paper I). Due to heavy foreground extinction ($A_{V} = 7.3$ mag), the galaxy is almost invisible in the optical, but the \HI\ contours demonstrate the notable extent of the disk ($\sim 130$ kpc). Near-infrared imaging of the galaxy reveals an early-type spiral (S0/Sa) with a prominent bulge, first noted by \citet{Don06}. However, the combined IRAC (Infrared Array Camera) bands, with a full width at half maximum (FWHM) of $\sim2$\arcs\ (see Fazio et al. 2004 for instrument details), show that in the mid-infrared the galaxy has a star forming disk approximately 50\,kpc in diameter. The extended star formation is best shown by the $[$Ne\,{\sc ii}$]$\,12.81\micron\ contours overlaid on a PAH three-colour image (FWHM $\sim2$\arcs) consisting of spectral maps of the 6.2, 7.7 and 11.3\micron\ emission bands. The PAH emission, excited in regions of star formation, extends into the disk where there is little emission from the old stellar population that dominates the near-infrared image; this suggests that the galaxy is building a disk from the inside out. The PAH map and $[$Ne\,{\sc ii}$]$\,12.81\micron\ contours reveal a clear asymmetry on the eastern side of the disk, a feature also seen in the 20-cm continuum contours and suggested by the \HI\ contours. The MIPS (Multiband Imaging Photometer; Rieke et al. 2004) 24\micron\ image (FWHM $\sim$6\arcs) resolves the galaxy, which appears as a point source at 70\micron\ (shown by the contours), and indicates vigorous star formation traced by dust reprocessing starlight. As discussed in Paper I, the combined photometric Spectral Energy Distribution (SED) is well-fitted by an Sc-type galaxy template with strong PAH and dust emission \citep[GRASIL code;][]{Sil98}.

In this paper we explore in detail the spectroscopic properties of HIZOA~J0836-43 using observations from the {\it Spitzer} Space Telescope and the Mopra mm-wave antenna. We build on the results of Paper I and provide new measurements of cold molecular gas (through CO observations on Mopra) to illustrate the seemingly unique (locally) nature of this system and discuss the importance of such a galaxy in the context of the star formation mechanism observed in disk galaxies at $z\sim1$.
HIZOA~J0836-43 is located behind the southern Milky Way (08$^{\rm h}$36$^{\rm m}$51.6$^{\rm s}$, -43$\degr$37\arcmin41.0\arcs) and we adopt a distance of $D_{L}$=148 Mpc, as found by \citet{Don06}, derived from its recessional velocity, $v_{hel}$=10 689 km\,$\rm{s^{-1}}$.

\section{Observations and Data Reduction}

The primary data set is from the {\it Spitzer Space Telescope} GO-3 programme 30914 and consists of imaging and spectroscopy centred on the galaxy. See Paper I for details of imaging observations.

\subsection{Spitzer Spectroscopy}

Spectroscopy of HIZOA~J0836-43 was obtained on 2007 May 2 and 3 using the IRS (Infrared Spectrograph) instrument \citep{Hou04} onboard {\it Spitzer}. Observations were done in low-resolution, $R\, (= \lambda/\Delta \lambda$) $\sim 64-128$, ``mapping mode" (short-low; SL) covering $5-14$\micron\ and consisting of $3 \times 60$\,s integrations. This set of observations consists of three separate mappings: Centre, East and West of the galaxy nucleus covering $0.4\arcmin \times 0.7\arcmin$ ($\sim \frac{1}{3}$ of the disk) with $\sim$10\% overlap. Outlier regions for background subtraction were observed with the same mapping scheme using a 26\arcs\ step parallel to the slit. 

High resolution ($R \sim650$) ``staring mode" observations were obtained of the nucleus, covering $10-20$\micron\ (short-high; SH) and $19-38$\micron\ (long-high; LH) with integrations of $4 \times 30$\,s and $4 \times 14$\,s, respectively. A region $\sim1\arcmin$ south of the galaxy without confusion from foreground Galactic emission was observed for background subtraction.

The {\it Spitzer} Science Center (SSC) IRS pipeline version S16.1.0. was used for spectral reductions such as wavelength and flux calibration, ramp fitting, dark subtraction and droop and linearity corrections. Basic Calibrated Data (BCDs) of the sky background (or ``outrigger'') observations for the SL Centre and West maps were used for background removal by means of the SSC tool, CUBISM\footnote{See http://ssc.spitzer.caltech.edu/archanaly/contributed/cubism} \citep{Smi07}. Contaminating emission in the East outrigger observation resulted in a combination of East, West and Centre outriggers being used for the East map background removal. This was done by smoothing the East background observations and subtracting them from the original frames to create an image of the East array's behaviour (with the background removed). The Centre and West outriggers were smoothed and combined to produce a background map. This was added to the East array map to make a suitable background for CUBISM. 
Outlier rejection was done using the CUBISM algorithm, as well as by visual inspection of the spectral cube. CUBISM was used to extract a spectrum of a $37^{\prime \prime}$ aperture, matched in SL1 ($1^{st}$ order) and SL2 ($2^{nd}$ order), to capture the infrared disk and nuclear region of the galaxy. SL2 was scaled to match SL1 in flux (scale factor of 1.48).

SH and LH data were reduced by subtracting median-combined background images (from the ``off" position observations) and carefully cleaning individual BCDs of bad pixels and cosmic ray contamination. CUBISM was used to combine the BCDs into a spectral cube and further outliers found by visual inspection.
A SH extraction of the nuclear region was performed using CUBISM ($\sim$ 9.25\arcs\ $\times$ 4.5\arcs) and a matched SL extraction performed. Due to differing slit sizes, the extracted LH aperture was slightly larger and the spectrum scaled down by 55\% to match the SH continuum flux at 19.3\micron.

Spectral features were measured using ISO Spectral Analysis Package (ISAP)\footnote{The ISO Spectral Analysis Package (ISAP) is a joint development by the LWS and SWS Instrument Teams and Data Centers. Contributing institutes are CESR, IAS, IPAC, MPE, RAL and SRON.}.

\subsection{Mopra CO(1-0) Observations}

Molecular line observations of HIZOA J0836--43 were carried out with the
22-m Mopra\footnote{The Mopra radio telescope is part of the Australia Telescope which is funded by the Commonwealth of Australia for operations as a National Facility managed by CSIRO.} telescope in the CO $J$(1-0) line (rest frequency 115.2 GHz)
during May 25--27, 2009 and August 12--18, 2010. The Mopra antenna is located near Coonabarabran
(NSW, Australia) at an altitude of 866 m and is part of the 
Australia Telescope National Facility (ATNF).
Our pointing position is $08{^{\rm h}}36{^{\rm m}}51.54{^{\rm s}}$, $-43\arcdeg37\arcmin$41.5\arcs, which corresponds to the 
20-cm radio continuum core of 
HIZOA J0836--43, as measured by Donley et al. (2006).
Using a 3\,mm receiver, the Mopra Spectrometer (MOPS) in wide-band mode covers a frequency range 
between 108.2 and 115.6\,GHz.  At the redshift of the galaxy, the CO line is centered at 111.3 GHz; each sub-band of MOPS is 2.2 GHz wide
with 8096 channels ($\times$ 2 polarisations) giving a channel width of 256.25 kHz (or 0.69 km/s).
At this frequency the antenna efficiency is 0.42 and the beam size is 32\arcsec.
With this beam size, approximately half of the star forming disk is covered; however, the molecular gas is expected to lie mostly in the nuclear and interior regions of the disk, largely covered by the Mopra beam.  Hence, we expect our observations to sample the bulk of the molecular gas in this system, but should be considered a lower limit. The galaxy was observed in an ``off1-on-on-off2'' position-switching mode, using two different ``off'' positions in order to minimize the deleterious effects of 30 MHz standing wave ripples that form between the dish and receiver. ``Off1'' is approximately 1\arcmin\ north of the galaxy (in Galactic coordinates) and ``off2'' approximately 1\arcmin\ south of the galaxy.  
A consistent calibration was maintained by measuring the system temperature (T$_{sys}$) every 10 to 15 minutes using a known-temperature blackbody ``paddle'' and adjusting T$_{sys}$ accordingly for the target observations. Typically, Tsys during our observations ranged between 250 and 400 K.

Combining data from the two observing runs, the galaxy was observed for a total of 16.6 on-source hours.  The basic observations were reduced using the ATNF-developed Python-based ATNF Spectral line Analysis package (ASAP), and scripts that were customised to our observation mode.  This produced calibrated spectra: $\nu$ (GHz) vs. antenna temperature (K). To remove the slowly undulating baselines, a Chebyshev polynomial was fit to the line-free channels and the baseline reassessed by measuring the residual RMS (Root Mean Square) and assigning a variance to the spectrum.  The typical baseline noise for an individual spectrum was $\sim$50\,mK. Spectra with particularly poor baseline noise, $>120$\,mK, due to the standing wave ripple were rejected. 
Once the antenna temperature was converted to a main-beam temperature, the individual observations and sub-bands were combined into one deep spectrum using an inverse-variance weighted-average of the spectra and performing outlier rejection. The final spectrum frequency was converted into a velocity assuming a redshift of $z=0.0356$. The achieved $1\sigma$ RMS sensitivity of the combined spectrum is 3.3\,mK at 111\,GHz. Applying Gaussian smoothing at 20\,km\,s$^{-1}$ resolution, comparable to the resolution of the Parkes \HI\ observations of the galaxy, the RMS sensitivity increases to 1.3\,mK. The flux density is related to the antenna temperature according to:
$S_{\nu} (Jy) = 2.65 \times (T_{a} / \eta) \times \theta [\arcmin]^2 / \lambda[cm]^2  $, where $\eta = 0.42$, $T_{a}$ is the antenna temperature, $\theta$ is the solid angle of the source and $\lambda$ is wavelength.

Additional observations centred on the southeastern disk warp feature (08$^{\rm h}$36$^{\rm m}$52.6$^{\rm s}$, -43$\degr$37\arcmin47.6\arcs), consisting of 2.2 on-source hours, achieved a smoothed RMS of 0.70\,mK (smoothing at 40\,km\,s$^{-1}$ resolution).

\vfill

\section{Results}\label{res}

\begin{figure*}[!tp]
\begin{center}
\includegraphics[width=12cm]{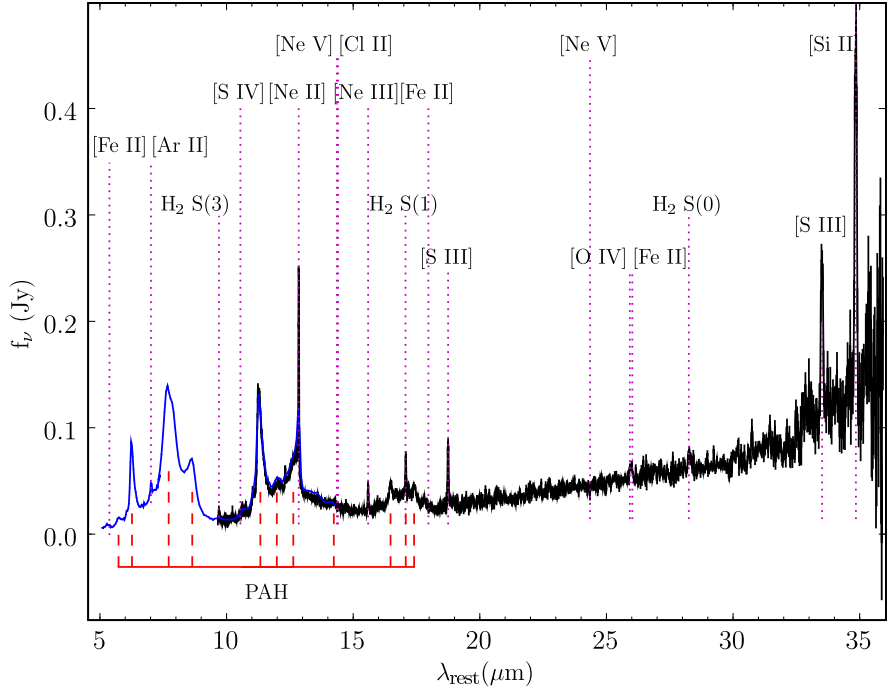}
\caption[Combined spectrum of HIZOA~J0836-43]{Combined low (SL; blue) and high resolution (SH, LH; black) spectrum of the nuclear region of HIZOA~J0836-43. Vertical lines indicate the locations of known emission lines (dashed purple lines), many of which are clearly detected, and prominent PAH lines (dashed red lines).}
\label{fig:spec_all}
\end{center}
\end{figure*}

\subsection{Spectroscopy of HIZOA~J0836-43}\label{spec}

Spectroscopy of HIZOA J08360-43 allows us to probe the conditions of the interstellar medium (ISM) and analyse the star formation mechanism. The larger SL mapping coverage allows us to study the emission in the nuclear region and the inner part of the star forming disk, but with only short wavelength coverage ($5-14$\micron) we can chiefly only probe the PAH emission. The measurements for the 37\arcs\ SL extraction are included and discussed in Appendix A.

We focus here on the combined spectrum (low and high resolution) of the nuclear region, presented in Fig.~\ref{fig:spec_all}. At wavelengths $< 15$\micron\ the galaxy's emission is dominated by PAH features. The excitation of these molecules is linked to large-scale star formation \citep{Gen98}; this is consistent with the LIRG photometric properties and SFR determined in \citet{Clu08}. Measurements of the PAH emission bands are listed in Table~\ref{tab:spec_ip}, and indicated in Fig. 2; they are discussed further in Section \ref{pah}.

In the high-resolution spectra of HIZOA~J0836-43, we detect the fine-structure lines of $[$Ar\,{\sc ii}$]$\,6.99\micron, $[$Ne\,{\sc ii}$]$\,12.81\micron, $[$Ne\,{\sc iii}$]$\,15.56\micron,  $[$S\,{\sc iii}$]$\,18.71\micron, $[$S\,{\sc iii}$]$\,33.38\micron\ and $[$Si\,{\sc iii}$]$\,34.82\micron. With the exception of the $[$Si\,{\sc ii}$]$\,34.82\micron\ line, these are all nebular lines from ionised hydrogen gas regions. The $[$Si\,{\sc ii}$]$ line, however, can be produced in ionised regions as well as warm atomic gas such as found in photodissociation regions (PDRs). The $[$O\,{\sc iv}$]$\,25.89\micron\ and $[$Fe\,{\sc ii}$]$\,25.99\micron\ lines are blended, but the absence of high-excitation emission in the spectrum makes $[$O\,{\sc iv}$]$\,25.89\micron\ unlikely. Similarly, we do not detect the high excitation $[$Ne\,{\sc v}$]$\,24.32\micron\ line, which in combination with the lack of $[$O\,{\sc iv}$]$\,25.89\micron\ emission, implies a weak, buried or absent Active Galactic Nucleus, or AGN, \citep{Stu02}. This agrees with the findings of \citet{Don06} based on their 20-cm continuum observations of HIZOA~J0836-43. The measured equivalents widths (EQW) and fluxes are shown in Table \ref{tab:spec_ip} and diagnostics associated with the detected lines are presented in Section \ref{ionic}.

\begin{table}[!thp]
{\scriptsize
\begin{minipage}{8cm}
\caption{PAH emission and Ionic line measurements. \label{tab:spec_ip}}
\begin{center}
\begin{tabular}{l r r l r  r}
\hline
\hline
\\[0.5pt]
Line     &  $\lambda_{rest} $   & FWHM     &  EQW   & $\nu F_{\nu}$ $^{a}$ & $\nu L_{\nu}$ $^{a}$  \\

       &    $\mu \rm{m}$       &  $10^{-3}\mu \rm{m}$    &    $  \mu \rm{m}$  &    $10^{-17}\,\rm{W}\, \rm{m}^{-2}$  & $10^{7} L_{\odot}$  \\
\hline
\\[0.5pt]
PAH  & 6.23  &   &   0.722   &     97.58 $\pm 2.14$   & 66.9 $\pm 1.5$
\\
PAH  & 7.67 &  &  0.819   &   219.21 $\pm 4.76$   &      150.0 $\pm 3.3$
\\
PAH  & 8.59  &  &  0.227   &   44.00  $\pm 1.52$     &   30.2 $\pm 1.0$
\\
PAH & 11.27 &  &   0.851  &  59.33  $\pm 1.03$      &   40.7 $\pm 0.7$
\\
PAH & 12.60  &  &  0.419   & 20.43  $\pm 1.37$       &   14.0 $\pm 0.9$
\\
PAH & 16.44  &  &   0.052   &  4.17  $\pm 0.47$       &   2.9 $\pm 0.3$
\\
PAH & 17.04  &  &   1.573   &   21.79 $\pm 2.25$      &   14.9 $\pm 1.5$
\\
PAH & 17.37  &  &   0.057    &   1.20  $\pm 0.41$         &   0.8 $\pm 0.3$
\\
$[$Ar\,{\sc ii}$]$   &  6.99$^{b}$  &   88.40   &   0.039    &    7.55 $\pm  0.62$ &   5.12 $\pm  0.11$   
\\
$[$Ne\,{\sc ii}$]$  &   12.81   &   32.80  &    0.208   &   13.02 $\pm  0.12$     &  8.92 $\pm  0.08$   
\\
$[$Ne\,{\sc iii}$]$    &    15.56    &    41.09    &     0.051          &  1.40 $\pm  0.08$ &  0.96 $\pm  0.06$   
\\
$[$S\,{\sc iii}$]$   &      18.71   &  52.40     &    0.165          &    3.18 $\pm  0.11$   &  2.12 $\pm  0.07$  
\\
$[$S\,{\sc iii}$]$    &      33.47   &    98.30       &    0.156  &    4.52 $\pm  0.28$     &  3.09 $\pm  0.20$    
\\
$[$Si\,{\sc ii}$]$    &      34.81   &     92.34      &  0.247   &  8.84 $\pm  0.48$   &   6.05 $\pm  0.33$     
\\
\hline
\\[0.5pt]

\hline
\\[0.5pt]
\multicolumn{6}{l}{$^{a}$ Error computed using the flux relative to the baseline RMS} \\
\multicolumn{6}{l}{$^{b}$ [Ar II] is blended with $\rm{H}_{2}$ {\it S}(2)}\\
\\

\end{tabular}
\end{center}
\end{minipage}
}
\end{table}

Conspicuously absent in the spectrum shown in Fig.~\ref{fig:spec_all} is a steeply rising warm dust continuum typically associated with a luminous infrared galaxy, 
where dust is heated due to copious star formation. Instead, it appears that the galaxy is dominated by cold dust, resulting in the spectrum with only a weak contribution from mid-infrared (MIR) continuum emission. Indeed the 6\micron\ and 15\micron\ continuum measurements are more indicative of a quiescent galaxy (Paper I). Very Small Grains (VSGs) are believed to radiate in the 15--60\micron\ wavelength region, i.e. between emission from PAHs and radiation from classical interstellar grains \citep{Pug89}. Ongoing studies explore the possibility that the VSGs are PAH ``clusters" and that isolated PAH molecules are actually produced by the destruction of these small carbonaceous grains \citep{Bou98,Rap06}. The spectrum of HIZOA~J0836-43 shows a distinctive lack of emission from these VSGs. This is seen quantitatively in continuum measurements; the flux densities at 15\micron\ and 30\micron\ are 0.022$\pm$0.002 Jy and 0.068$\pm$0.004 Jy respectively. In their sample of starburst systems, \citet{Br06} determined a relation between the total infrared luminosity ($L_{\rm{TIR}}$) of a galaxy and the power in its continuum. The values obtained for HIZOA~J0836-43 (above) correspond to $L^{\rm{est}}_{\rm{IR}}({\rm continuum}) \approx 1.8 \times 10^{10} L_{\odot}$. This is $\sim$ 6 times smaller than the value determined from the broadband photometry in Paper I ($L_{\rm{TIR}} \sim 1.1 \times 10^{11} L_{\odot}$). This illustrates the relative weakness of the continuum in the nuclear region in comparison to the total infrared emission from the galaxy.


\subsection{PAH Emission Line Diagnostics}\label{pah}

Measurements of the PAH features of HIZOA~J0836-43, listed in Table \ref{tab:spec_ip} and indicated in Fig. 2, have been made using a spline continuum as described by \citet{Peet02} and \citet{Sp07}. For completeness (and further comparison) we include measurements made using the PAH decomposition method \citep[PAHFIT; as described by][]{Smi07} in Appendix B.

The PAH emission in HIZOA~J0836-43 appears uncommonly strong, relative to the continuum, with a 7.7\micron\ equivalent width of 0.82\micron\ compared to the average of 0.53\micron\ for the starburst sample of \citet{Br06}. The galaxy has a 7.7\micron/11.3\micron\ PAH flux ratio of $3.69 \pm 0.11$, which is a factor of $\sim2$ greater than that found for typical starbursts. For the 8.6\micron\ and 11.3\micron\ PAH feature, \citet{Br06} find a range of EQWs from $0.003-0.192$\micron\ and $0.180-0.931$\micron\ respectively. HIZOA~J0836-43 has corresponding EQWs of 0.227\micron\ and 0.851\micron\ by comparison. However, it is the 17\micron\ PAH feature which clearly shows the difference between HIZOA~J0836-43 and other starbursts. The largest value is an EQW of 0.792\micron\ for the galaxy NGC 7252, while in strong contrast, HIZOA~J0836-43 has an EQW of 1.573\micron, almost twice as large as that found for NGC 7252. We attribute these large equivalent widths to a combination of power in the PAH emission lines and the absence of a sharply rising dust continuum typical of starburst systems.

PAH emission is closely related to the total infrared luminosity of a starbursting system since they both measure star formation activity. The 6.2\micron\ PAH flux should scale with the total infrared luminosity of the system within a factor of $\sim2$ \citep{Br06}. Using the relation of \citet{Br06} derived from their starburst sample, and applying it HIZOA~J0836-43, the estimated infrared luminosity is $L^{\rm{IR}}_{\rm{est}}({\rm PAH}) \approx 4.7 \times 10^{11} L_{\odot}$. This metric overestimates by a factor of $\sim$ 4 the total luminosity determined from photometry. This further illustrates the strong (nuclear) PAH emission in relation to the total infrared emission of the galaxy, most notably the weak emission from a warm dust continuum. Strong PAH emission coupled with similarly strong continuum emission would result in a large total infrared luminosity. HIZOA~J0836-43 appears atypical, as it has powerful PAH emission, but lacks similarly strong continuum emission, thus giving rise to the observed large discrepancies in the star-formation metrics as compared to other starburst systems. 


\citet{Br06} find a relation between $L(6.2\mu m)$ PAH and $\nu L_{\nu}(5.5\mu m)$ for low-luminosity starbursts, $ {\rm{log}}[L(6.2\mu m) {\rm PAH}] = -0.49 +0.96\, {\rm{log}} [ \nu L_{\nu}(5.5\mu m)]$. Substituting the $\nu L_{\nu}(5.5\mu m)= 6.35 \times 10^{9} L_{\odot}$ yields a value of $L(6.2\mu m) = 8.3 \times 10^{8} L_{\odot}$, close to the measured value of $L(6.2\mu m) = 6.7 \times 10^{8} L_{\odot}$. HIZOA~J0836-43 therefore lies on the correlation, but with a luminosity greater than that seen in local starbursts, but less than that seen in ULIRGs (Ultra Luminous Infrared Galaxies) before they ``turn off'' the relation; see Desai et al. 2007.


The total power in the PAH emission compared to the total infrared luminosity of the galaxy can be illustrated by summing over the luminosities of the PAH emission features (see Table \ref{tab:spec_ip}), yielding a total PAH luminosity of $L_{\rm{PAH}} = 3.2 \times 10^{9} L_{\odot}$. Given that the galaxy has a total infrared luminosity of $L_{\rm{TIR}} = 1.2 \times 10^{11} L_{\odot}$, the nuclear PAH emission constitutes $\sim 2.7\%$ of the total infrared emission. The 10 ULIRGs studied by \citet{Ar07} have percentages ranging from $0.4-2.1$\% by comparison. In fact, we show in section 5 (see also Fig. 4), only hyper-ULIRGs have PAHs of similar strength (relatively), compared to HIZOA~J0836-43.

\subsection{Ionic Emission Line Diagnostics}\label{ionic}

In the following analysis we use the relative strengths of the emission lines (presented in Table \ref{tab:spec_ip}) to probe the heating mechanisms, metallicity, electron density and star formation activity in the central region of HIZOA~J0836-43.

The ratios of the fine structure line fluxes of different ionic species of the same element (eg. $\rm{Ne^{2+}}/\rm{Ne^{1+}}$) provides a measure of the excitation and thus an indication of the hardness of the radiation field. 
The ratio of $[$Ne\,{\sc iii}$]$\,15.56\micron/$[$Ne\,{\sc ii}$]$\,12.81\micron\ (ionization potential of 41 eV/21.6 eV) is $\sim$ 0.11, comparable to typical values for starbursting systems \citep{Br06,Ver03}.
Although the $[$Ne\,{\sc iii}$]$/$[$Ne\,{\sc ii}$]$ ratio is not uncommon, it is below the median value (0.28) found for star forming regions in the SINGS ({\it Spitzer} Infrared Nearby Galaxy Survey) sample \citep{Dale09}. This implies greater weight in the $[$Ne\,{\sc ii}$]$\,12.81\micron\ flux compared to $[$Ne\,{\sc iii}$]$\,15.56\micron. The $[$Ne\,{\sc ii}$]$\,12.81\micron\ emission acts as an indication of the surface density of \HII\ regions as it has a much lower ionisation potential compared to $[$Ne\,{\sc iii}$]$\,15.56\micron. The low $[$Ne\,{\sc iii}$]$/$[$Ne\,{\sc ii}$]$ ratio indicates a relatively soft radiation field resulting in \HII\ emission and potentially large amounts of emission from PDRs.


The ratio of $[$S\,{\sc iii}$]$\,18.71\micron\ to $[$S\,{\sc iii}$]$\,33.48\micron\ is an electron density diagnostic. The $[$S\,{\sc iii}$]\, 18.71$\micron/$[$S\,{\sc iii}$]\, 33.48$\micron\ ratio of $\sim$ 0.70 corresponds to an electron density of $n_{e} \sim 300\, \rm{cm}^{-3}$ \citep{Smi09}, typical of starbursting and normal disk systems \citep{Ver03,Dale09}.

Dense PDRs and X-ray dominated regions, powered by AGN, show strong $[$Si\,{\sc ii}$]$ 34.82\micron\ emission due to the significant cooling effect of this line \citep{Holl99}, while the $[$S\,{\sc iii}$]$ 33.48\micron\ line acts as a strong tracer of \HII\ regions \citep{Dal06}.
Compared to the AGN and star forming regions in \citet{Dale09}, the $[$Si\,{\sc ii}$]$\,34.82\micron/$[$S\,{\sc iii}$]$\,33.48\micron\ ratio of $1.96 \pm 0.16$ is high compared to star forming regions in SINGS, but toward the low-end for the AGN distribution. Given the lack of an AGN tracer in its spectrum, it appears that HIZOA~J0836-43 has large amounts of dense PDRs producing powerful $[$Si\,{\sc ii}$]$ emission which dominates the $[$S\,{\sc iii}$]$ coming from \HII\ regions, causing the inflated $[$Si\,{\sc ii}$]$\,34.82\micron/$[$S\,{\sc iii}$]$\,33.48\micron\ ratio. Dust grain destruction, such as caused by an AGN or shocks, will increase the amount of $[$Si\,{\sc ii}$]$\,34.82\micron\ and $[$Fe\,{\sc ii}$]$\,25.99\micron\ in the interstellar medium \citep{Dale09}. These species are therefore enhanced in active galaxies. In HIZOA~J0836-43 we do not detect significant $[$Fe\,{\sc ii}$]$\,25.99\micron\, which indicates that $[$Si\,{\sc ii}$]$\,34.82\micron\ enhancement is coming from elsewhere. We also know that the radiation field is not hard enough to encourage such a mechanism. Since $[$Si\,{\sc ii}$]$\,34.82\micron\ traces strong PDR emission, we conclude that the elevated $[$Si\,{\sc ii}$]$\,34.82\micron/$[$S\,{\sc iii}$]$\,33.48\micron\ ratio is the result of strong PDR emission compared to emission from \HII\ regions. 
We summarise in Table \ref{tab:spec_prop} the main properties from Sections 3.2 and 3.3.

\begin{table}[!t]
{\scriptsize
\begin{minipage}{8cm}
\caption{Summary of spectroscopic and derived properties of HIZOA~J0836-43\label{tab:spec_prop}}
\begin{center}
\begin{tabular}{l l }
\hline
\hline
  Property      &  Value \\[0.5pt]
\hline
\\
6.2\micron\ PAH EQW     & 0.72\micron
\\
7.7\micron\ PAH EQW    &  0.82\micron
\\
${\rm F}(7.7 \mu {\rm m\, PAH})/ {\rm F}(11.3 \mu {\rm m\, PAH})$        &    3.69 $\pm 0.10$
\\

$L_{\rm{PAH}} $  &   $ 3.2 \times 10^{9} L_{\odot}$

\\
$[$Ne\,{\sc iii}$]\,15.56$\micron/$[$Ne\,{\sc ii}$]\,12.81$\micron\  &  0.11 $\pm0.01$
\\
$[$S\,{\sc iii}$]\,18.71$\micron/$[$S\,{\sc iii}$]\,33.48$\micron\ &  0.70 $\pm0.05$
\\
$[$Si\,{\sc ii}$]\,34.82$\micron/$[$S\,{\sc iii}$]\,33.48$\micron   &  1.96 $\pm0.16$
\\
$n_{e}$   $^{a}$  &   $\sim300 \, \rm{cm^{-3}}$
\\

$L_{\rm{est}}^{\rm{IR}}({\rm PAH})$ $^{b}$  &    $\approx 4.7 \times 10^{11} L_{\odot}$
\\
$L_{\rm{est}}^{\rm{IR}}({\rm Continuum}) $ $^{c}$  &   $\approx 1.8 \times 10^{10} L_{\odot}$
\\
$L_{\rm{TIR}} $ $^{d}$  &   $1.2 \times 10^{11} L_{\odot}$
\\
\hline
\\[0.5pt]
\multicolumn{2}{l}{$^{a}$ Estimated from $[$S\,{\sc iii}$]$18.71\micron/$[$S\,{\sc iii}$]$33.48\micron\ ratio} \\
\multicolumn{2}{l}{$^{b}$ Estimated from 6.2\micron\ PAH Flux } \\
\multicolumn{2}{l}{$^{c}$ Estimated from 15\micron\ and 30\micron\ flux densities}\\
\multicolumn{2}{l}{$^{d}$ Cluver et al. (2008)}\\

\hline
\\[0.5pt]
\\
\end{tabular}
\end{center}
\end{minipage}
}
\end{table}


\subsection{Star Formation Rates}

In this section we use several metrics to estimate the star formation in HIZOA~J0836-43. Given the galaxy's unusual properties, we expect a range of values depending on the metric used. We have seen that the PAH features appear relatively strong compared to the warm dust continuum and we expect high SFRs from PAH metrics and low SFRs from warm dust continuum metrics. Since the SFR relations have not been calibrated using galaxies such as HIZOA~J0836-43, most should be interpreted qualitatively. Our relatively conservative number of $20.5\, M_{\odot}\, \rm{yr^{-1}}$, determined from the FIR emission, provides a credible measure of the star formation through dust reprocessed starlight; we retain this as our principal SFR for comparisons.

The strongest PAH features are useful measures of SFR, since the mechanism of their production and excitation is closely linked to star formation. \citet{Far07} used the scaling between $[$Ne {\sc ii}$]$ + $[$Ne {\sc iii}$]$ and the combined 6.2\micron\ and 11.3\micron\ luminosities for their sample of 57 ULIRGs, to obtain the relation ${\rm SFR}(M_{\odot}\, {\rm yr^{-1}}) = 1.18 \times 10^{-41} L_{P} (\rm{erg\, s^{-1}})$. Substituting HIZOA~J0836-43's value of $L_{P} = 4.12 \times 10^{42}\, \rm{erg\, s^{-1}}$ predicts a SFR of $\sim 49\, M_{\odot}\, \rm{yr^{-1}}$, approximately twice the SFR calculated using the FIR photometry. This suggests a relative excess of PAH emission in comparison to the total IR emission of the galaxy, already seen in Section \ref{pah}. It is, however, closer to the 20-cm radio continuum derived quantity ($\sim 35\, M_{\odot}\, \rm{yr^{-1}}$) from \citet{Don06}, assuming that star formation dominates this emission.

\citet{Hou07} use the starburst sample of \citet{Br06} to derive a SFR relation based on the scaling between $\nu L_{\nu}$(7.7\micron) and $L_{\rm IR}$, yielding log[SFR] ($M_{\odot}\, \rm{yr^{-1}}$) = log[$\nu L_{\nu}$(7.7\micron) erg\,s$^{-1}$] $-42.57$. Applying the 7.7\micron\ luminosity from Table 1 gives a SFR (in the nuclear region) of $24.7 M_{\odot}\, \rm{yr^{-1}}$.

The sum of the $[$Ne\,{\sc ii}$]$ and $[$Ne\,{\sc iii}$]$ luminosities is tightly correlated with total infrared luminosity, which in turn is correlated to the star formation rate \citep{Ho07}. This relation was modified by \citet{Far07}, obtaining:
\[ {\rm SFR}(M_{\odot}\, {\rm yr^{-1}}) = 2.69 \times 10^{-41} \frac{L_{N} (\rm{erg\, s^{-1}})}{f_{ion}(f_{\rm Ne^{+}} + 1.67f_{\rm Ne^{++}})} \ \ \ \ \ \ \]

\noindent where $L_{N}$ is the combined luminosity of the two neon lines, $f_{ion}$ is the fraction of photons contributing to the ionising the gas and $f_{\rm Ne^{+}}$ and $f_{\rm Ne^{++}}$ are the fractional abundances of $[$Ne\,{\sc ii}$]$ and $[$Ne\,{\sc iii}$]$ respectively. Using $f_{ion} =0.6$, $f_{\rm Ne^{+}} = 0.75$ and $f_{\rm Ne^{++}} = 0.1$ \citep{Ho07} and $L_{N} = 3.95 \times 10^{41}\, \rm{erg \, s^{-1}}$, yields a SFR of $\sim 20.8 M_{\odot}\, {\rm yr^{-1}}$. 

This line emission derived SFR agrees well with the SFR obtained using the FIR photometry, $20.5\, M_{\odot}\, \rm{yr^{-1}}$, indicating that the star formation is consistent with the strength of the ultraviolet radiation field.

The correspondence between SFR and total infrared luminosity (TIR) for infrared galaxies has been investigated by \citet{Riek09} with SFR equations derived using the MIPS $L_{24}$ measurements. They find SFR = $7.8 \times 10^{-10}L_{24}(7.76 \times 10^{-11}L_{24})^{0.048}$ where $L_{24}$ is in $L_{\odot}$ and for galaxies with $L_{\rm{TIR}} > 10^{10} L_{\odot}$. HIZOA~J0836-43 would have a SFR of $8.5 M_{\odot}\, \rm{yr^{-1}}$ using this metric. Since the 24\micron\ dust continuum appears weak in the spectrum of HIZOA~J0836-43, we would expect this to underestimate the total star formation in the galaxy.

We have SL coverage ($5-14$\micron) for the nuclear region and the inner MIR disk, so we can use the larger (37\arcs) extraction area to probe the PAH emission in the disk of HIZOA~J0836-43 (Appendix A) and use these values to predict the SFR. Strong PAH emission from the star forming regions in the disk results in high luminosities (see Table 5). Using the 7.7\micron\ PAH value in the relation of \citet{Hou07} (see above) yields a SFR of $47.4 M_{\odot}\, \rm{yr^{-1}}$. The 6.2\micron\ and 11.3\micron\ PAH luminosities in the relation of \citet{Far07}, see above, predicts a SFR of $134.5M_{\odot}\, \rm{yr^{-1}}$. The extended star formation in the disk of HIZOA~J0836-43 and the resulting PAH emission suggests a remarkably high SFR. We summarise the SFRs mentioned here in Table \ref{tab:sfr}.

\begin{table}[!htbp]
{\scriptsize
\begin{minipage}{8cm}
\caption{Derived Star Formation Rates for HIZOA~J0836-43\label{tab:sfr}}
\begin{center}
\begin{tabular}{l l l}
\hline
\hline
  SFR &  Input &  Reference \\[0.5pt]
\hline
\\
  $49\, M_{\odot}\, \rm{yr^{-1}}$  & 6.2\micron, 11.3\micron\ PAH  $^{a}$ & Farrah et al. (2007)
\\
  $ 24.7\, M_{\odot}\, \rm{yr^{-1}}$  & 7.7\micron\ PAH  $^{a}$ & Houck et al. (2007)
\\
 $ 21\, M_{\odot}\, {\rm yr^{-1}}$   & $[$Ne\,{\sc iii}$]$/$[$Ne\,{\sc ii}$]$ & Farrah et al. (2007)
\\
 $ 8.5\, M_{\odot}\, {\rm yr^{-1}}$  & 24\micron\ & Reike et al. (2009)
\\
 $ 47.4\, M_{\odot}\, {\rm yr^{-1}}$    &  7.7\micron\ PAH  $^{b}$ & Houck et al. (2007)
\\
 $ 134.5\, M_{\odot}\, {\rm yr^{-1}}$  & 6.2\micron, 11.3\micron\ PAH  $^{b}$ & Farrah et al. (2007)
\\
 $ 35\, M_{\odot}\, {\rm yr^{-1}}$ $^{c}$ & 20-cm continuum  &  Bell (2003)
\\
$ 20.5\, M_{\odot}\, {\rm yr^{-1}}$ $^{d}$ & FIR photometry & Kennicutt (1998)
\\
\hline
\\[0.5pt]
\multicolumn{3}{l}{$^{a}$ Nuclear extraction}\\
\multicolumn{3}{l}{$^{b}$ 37\arcs\ extraction}\\
\multicolumn{3}{l}{$^{c}$ Estimated from 20-cm continuum; Donley et al. (2006)}\\
\multicolumn{3}{l}{$^{d}$ Estimated from FIR photometry; Cluver et al. (2008)}\\

\hline
\\[0.5pt]
\\
\end{tabular}
\end{center}
\end{minipage}
}
\end{table}


\section{Molecular Gas}\label{molhy}

\subsection{Warm Molecular Hydrogen}


Emission lines from pure rotational excitation of molecular hydrogen (S(0), S(1), S(2) and S(3) transitions) are also observed in the spectrum of the galaxy (Fig. \ref{fig:spec_all}) and likely arise in PDRs neighbouring H {\sc ii} regions, but other possible heating mechanisms are shock heating, X-ray heating and cosmic ray heating \citep{Rig02}. These emission lines act as a powerful probe of the interstellar medium as they constrain the energy injection that excites the $\rm{H_{2}}$. The measured properties of these lines for HIZOA~J0836-43 are summarised in Table \ref{tab:spec_h2} and their excitation can be used to estimate the temperature and mass of (warm) molecular hydrogen. We obtain a mass of $1.3\times 10^{7}$ M$_\sun$ and a temperature of 333\,K.





Although this appears to be a relatively ``typical" amount of warm $\rm{H_{2}}$ (albeit on the low side) for a starburst galaxy \citep{Rig02}, it is only 0.02\% of the total \HI\ gas in the galaxy. There exists an apparent relationship between the luminosities of the 7.7\micron\ PAH and $\rm{H}_{2}$ S(1) emission lines \citep{Rig02}, since both types of emission are thought to originate in molecular clouds with star formation. The measured values for HIZOA~J0836-43 are $L_{7.7 \mu {\rm m}} = 5.8 \times 10^{35}\, \rm{W}$ and $L_{\rm{H_{2} S(1)}} = 3.3 \times 10^{32}\, \rm{W}$. The starbursts in the sample of \citet{Rig02} appear to follow the relation: log $L_{7.7 \mu {\rm m}}$ = 11.01 + 0.69 log $L_{{\rm H}_2 S(1)}$. Substituting $L_{7.7 \mu {\rm m}}$ implies $L_{\rm{H_{2} S(1)}} \approx 7.5 \times 10^{35}\, \rm{W}$, three orders of magnitude larger than what is observed. This suggests a relative paucity of warm $\rm{H_{2}}$ given the LIRG starburst in HIZOA~J0836-43. This combined with the exceptionally strong PAH and distinctively weak MIR continuum suggests atypical interstellar medium conditions or heating mechanisms compared to the local starburst galaxy population.

\begin{table}[!htbp]
{\scriptsize
\begin{minipage}{8cm}
\caption{$\rm{H_{2}}$ Emission Line Measurements (Nucleus) \label{tab:spec_h2}}
\begin{center}
\begin{tabular}{l c c c c}
\hline
\hline
\\[0.5pt]
  Line     &  $\lambda_{rest} $   & FWHM     &  EQW   & Flux  \\

       &    $\mu \rm{m}$       &  $10^{-3}\mu \rm{m}$    &    $ \mu \rm{m}$ &  $10^{-17} \ \rm{W} \ \rm{m}^{-2}$ \\
\hline
\\
$\rm{H}_{2}$ {\it S}(3) &   9.66    &     32.66  &   0.037   &      1.31  
\\
$\rm{H}_{2}$ {\it S}(2) $^{a}$ &  12.28   &   24.63   &   0.009   &   0.920   
\\
$\rm{H}_{2}$ {\it S}(1) &   17.04    &    41.0   &   0.186 &  1.57   
\\
$\rm{H}_{2}$ {\it S}(0) &   28.22    &   98.0   &   0.042    &     0.61  
\\
\hline
\\[0.5pt]
\multicolumn{5}{l}{$^{a}$ $\rm{H_{2}}$ {\it S}(2) is blended with [Ar II]} \\
\\
\end{tabular}
\end{center}
\end{minipage}
}
\end{table}

\subsection{Molecular CO}

The interplay between neutral and molecular hydrogen in HIZOA~J0836-43 is of primary importance to understanding the starburst mechanism and interstellar medium conditions we observe. We therefore need to gauge the amount of cold H$_2$ as traced by CO gas. This key observable allows us to probe the efficiency of star formation, as well as compare to other star forming systems.

\begin{figure}[!tbh]
\begin{center}
\includegraphics[width=8cm]{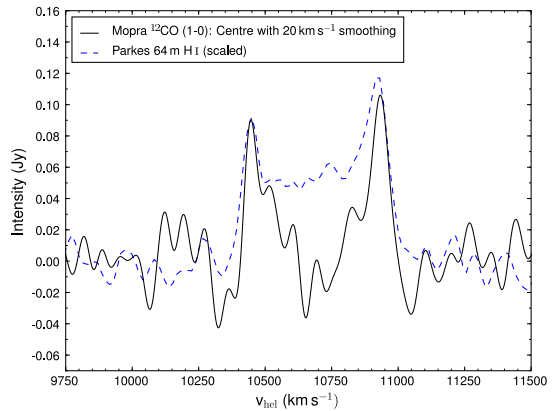}
\caption{$^{12}$CO $J(1-0)$ Mopra detection corresponding to 16.6 on-source hours after applying Gaussian smoothing of 20\,km\,s$^{-1}$ (solid line); the achieved $1\sigma$ sensitivity of the combined spectrum is 0.8\,mK. For comparison the \HI\ profile from Parkes \citep{Don06}, scaled to match the CO peak at 10\,400\,km\,s$^{-1}$, is shown as the blue dashed line.}
\label{fig:cold}
\end{center}
\end{figure}

The relatively large distance of HIZOA J0836-42, 148 Mpc, combined with the exceptionally broad velocity width of the CO\,(1-0) line presents a serious challenge in detecting the molecular gas with the single 22\,m dish Mopra telescope -- the line is mostly lost in the noise at $\sim$5\,mK sensitivity with 1 km\,s$^{-1}$ resolution.  However, if we smooth to a scale of $\sim$20 km\,s$^{-1}$, which is close to the optimal bin size for the peaks in the \HI\ rotation curve (and the resolution of the Parkes \HI\ observations), then we clearly detect the CO emission (see Fig. \ref{fig:cold}). The molecular gas has a rotation curve that is nearly identical to that of the atomic gas, at least for the velocity peaks where we have the best signal-to-noise, spanning a velocity range of $\pm$ 315\,km\,s$^{-1}$.

At low rotational velocities ($\pm 100$\,km\,s$^{-1}$), the CO gas is not detected, lying just below the 1 to 2$\sigma$ noise limit. Integrating across the entire 630\,km\,s$^{-1}$ line width, the velocity-integrated intensity is 1.59\,K\,km\,s$^{-1}$ (16.5\,Jy\,km\,s$^{-1}$), achieved with a S/N (signal to noise) of 3.3.   Using these values, the total molecular gas can be estimated. Assuming the line is optically thin and a standard Galactic conversion factor of N(H$_{2}$)/ I$_{\rm CO}$ = $2\times10^{20}$\,cm$^{-2}$, Lisenfeld et al. (2002) showed that the total molecular gas mass is M(H$_2$)[M$_\sun$] = 75 I$_{\rm CO}$ $D^2$ $\Omega$, where $D$ is in Mpc and $\Omega$ is the area in arcsec$^2$.  Using this formula, we find that the H$_{2}$ column density is $3.18\times10^{20}$\,cm$^{-2}$, and the total molecular gas mass (H$_{2}$ + He, where the He fraction is 1.38) is 3.7$\times 10^{9}$ M$_\sun$. By comparison the relations of \citet{Sol92} and \citet{Evans99}, using the Milky Way conversion factor ($\alpha=4.6$), both yield $3.9 \times 10^{9}$\,M$_\sun$.

\subsection{Gas Diagnostics}

Comparing the total molecular gas mass to the total atomic hydrogen mass, the H$_{2}$ to \HI\ gas-mass fraction is $\sim$5\%. We do not detect the low velocity gas seen in the \HI\ distribution (see Fig. \ref{fig:cold}) and given the achieved sensitivity of our observations we find it unlikely that it is present. If this galaxy behaves like most gas-rich spiral galaxies observed in the local universe, we expect that most of the gas will be located within our beam. However, to determine a complete census of the molecular gas in the galaxy, we need to fully map the entire star forming disk ($\sim$1\arcmin\ vs. 30\arcsec\ Mopra beam), and in particular the gas content in the ``warp'' region. Additional observations centred on the southeastern disk warp feature (08$^{\rm h}$36$^{\rm m}$52.6$^{\rm s}$, -43$\degr$37\arcmin47.6\arcs), consisting of only 2.2 on-source hours, resulted in a weak detection of molecular gas with a S/N of $\sim$ 2 (for the $-300$\,km\,s$^{-1}$ velocity peak). This hints at the molecular gas reservoir being extended in the disk, but additional observations are required to verify the distribution, and obtain a complete inventory, of the molecular gas content. We, therefore, treat the value obtained above as corresponding to a lower limit for the total molecular gas in the disk and adopt a H$_{2}$ column density of $3.18 \times 10^{20}$\,cm$^{-2}$, corresponding to a H$_{2}$ + He mass of $3.7 \times 10^9$ M$_\sun$ (using $\alpha=4.6$), for all further comparisons in this paper.

As shown by \citet{Gao04}, the far-infrared luminosity appears correlated with the amount of molecular gas in LIRGS. This formula applied to HIZOA~J0836-43 predicts $1.3 \times 10^{10}$ M$_\sun$ of H$_{2}$, approximately a factor of three higher than what is implied from CO observations, supporting the possibility that our detection is missing a significant fraction of molecular gas. 

The cold gas fraction, $f_{\rm{cold}} = M{\rm(H_{2})}/[M{\rm(H_{2})}+M_{*}]$, can be used to estimate the progress of stellar building. In HIZOA~J0836-43 this fraction corresponds to $\sim$8\% of cold gas, which appears low compared to the local LIRG sample of \citet{Wang06} given the galaxy's stellar mass. 
This could be due to our CO measurement not capturing the extended component of the gas, however, HIZOA~J0836-43 does not appear to be a typical local LIRG and we compare to these systems with caution.

\section{Interpretation of Observations}\label{mir_prop}



HIZOA~J0836-43 appears to have strong PAH emission, but weak dust continuum emission. As shown in Section \ref{res} the PAH luminosity overpredicts the total infrared luminosity ($L_{\rm{TIR}} = 1.2 \times 10^{11} L_{\sun}$) of the galaxy by a factor of $\sim 4$. Conversely, the continuum fluxes underestimate $L_{\rm{TIR}}$ by a factor of $\sim 6$. To illustrate this discrepancy we calculate the (synthetic) integrated 24\micron\ emission using the spectrum of the nuclear region of HIZOA~J0836-43 (Fig. 2) and compare that to the luminosity seen in the PAH bands. The spectral flux of 0.126 Jy corresponds to a luminosity of $L_{24}=2.4 \times 10^{9}L_{\sun}$. Using the total PAH power from Table 1 ($L_{\rm PAH}=3.2 \times 10^{9}L_{\sun}$) we find a ratio of $L_{\rm PAH}$/$L_{24}$ in the nuclear region of $\sim$ 1.3. If we use the PAH luminosities measured using PAHFIT (Smith et al. 2007) given in Appendix B, $L_{\rm PAH}=9.7 \times 10^{9}L_{\sun}$ and $L_{\rm PAH}$/$L_{24} \sim 4.0$.

\begin{figure*}[!thbp]
\begin{center}
\includegraphics[width=14cm]{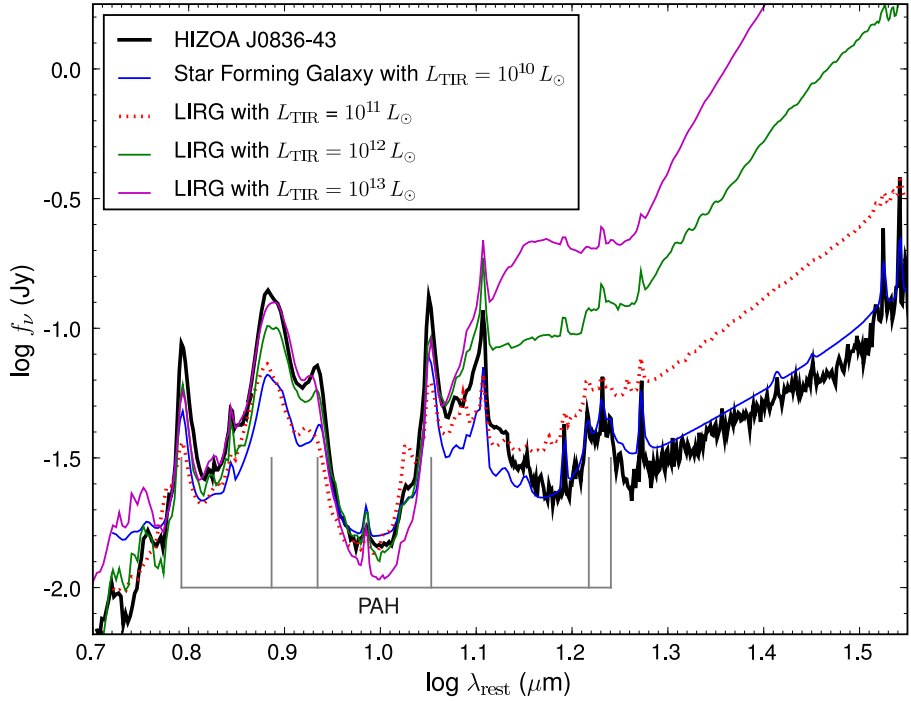}
\caption{{\it Spitzer} spectrum of HIZOA~J0836-43 (thick, black line) with scaled SED templates from Rieke et al. (2009) for starforming galaxies with $L_{\rm TIR}$=10$^{10}L_{\sun}$ (blue, solid), $L_{\rm TIR}$=10$^{11}L_{\sun}$ (red, dotted) and $L_{\rm TIR}$=10$^{12}L_{\sun}$ (green, solid) and $L_{\rm TIR}$=10$^{13}L_{\sun}$ (magenta, solid. The relative PAH strength in HIZOA~J0836-43 significantly exceeds that seen in LIRGs of similar infrared luminosity.}
\label{fig:lirg}
\end{center}
\end{figure*}

The unusual combination of strong PAH emission relative to a weak warm dust continuum is shown graphically in Figure \ref{fig:lirg}. This plots the SED templates for luminous and ultraluminous, purely star forming infrared galaxies, from \citet{Riek09}, scaled to match the spectrum of HIZOA~J0836-43 (using the short wavelength continuum). The warm dust continuum ($\lambda > 20$\micron\ of the galaxy is best fit by a star forming galaxy with $L_{\rm TIR}$=10$^{10}L_{\sun}$, however, the PAH stength is more similar to a hyper-ULIRG with $L_{\rm TIR}$=10$^{13}L_{\sun}$. Compared to LIRGs with similar infrared luminosity ($L_{\rm TIR}$=10$^{11}L_{\sun}$), the 6.2\micron\ and 7.7\micron\ PAHs in HIZOA~J0836-43 are $\sim$2 times stronger and the 11.3\micron\ PAH is 
 $\sim$2.5 times more luminous.

\begin{figure}[!thb]
\begin{center}
\epsfxsize=8cm
\epsffile{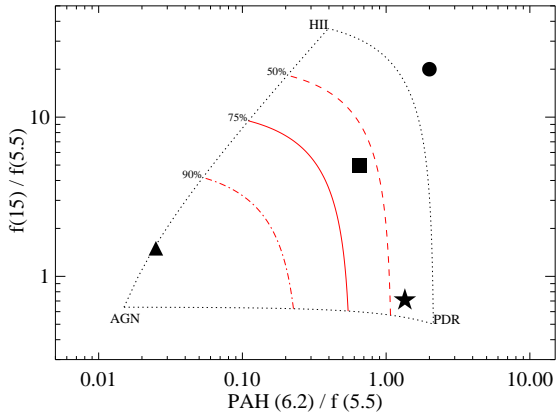}
\caption[MIR diagnostic diagram from Armus et al. (2007)]{MIR diagnostic diagram from \citet{Ar07}; the three vertices represent prototypical AGN, \HII\ and PDR emission, respectively, and the dashed lines show the fractional AGN contribution to the MIR emission. HIZOA~J0836-43 is shown (star), as well as the ULIRGs Markarian 231 (triangle) and Arp 220 (square), and the starburst galaxy NGC 3256 (circle) for comparison (data points from Armus et al. 2007).}
\label{fig:ar}
\end{center}
\end{figure}

PAH strengths and continuum fluxes can be used to determine the dominant mechanisms producing MIR emission. \citet{Ar07} modified the diagnostic of \citet{Laur00} to determine the relative contributions of MIR emission sources in ULIRGS, AGN and starburst galaxies. Using the 15 to 5.5 \micron\ continuum flux ratio (to separate starbursts and AGN) versus the 6.2 \micron\ equivalent width to the 5.5 \micron\ continuum (to determine the quiescent star formation contribution), one can discern between AGN, \HII\ and PDR-dominated emission. 
Using this diagnostic (shown in Fig. \ref{fig:ar}), HIZOA~J0836-43's MIR emission appears strongly dominated by emission arising from PDRs to the extent that emission from AGN and \HII\ regions is virtually insignificant. The relative strength of the PAH bands compared to the continuum emission in HIZOA~J0836-43 is amongst the largest observed in any star-forming galaxy, implying unusually strong PDR-dominated emission. It appears similar to the relative emission determined for the reflection nebulae, NGC 7023 and NGC 2023 \citep{Peet04}, but with stronger 6.2\micron\ PAH emission. Both are considered prototypical of PDR emission, arising from B stars producing a small region of ionised gas and illuminating a shell of molecular gas \citep{Wyr00,Fuen00}.

\begin{figure}[!thbp]
\begin{center}
\includegraphics[width=8cm]{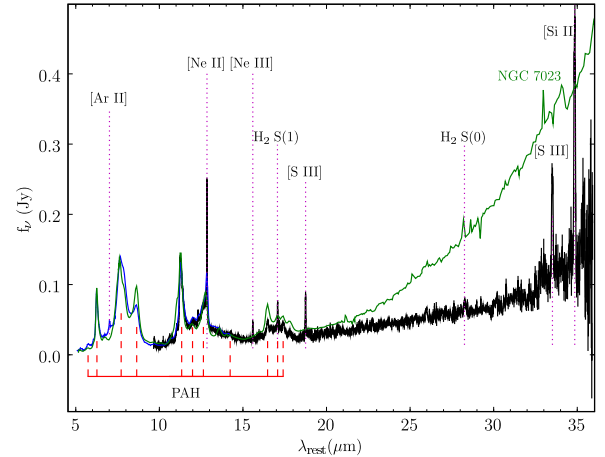}
\caption[Spectrum of HIZOA~J0836-43 compared to NGC 7023]{The spectrum of HIZOA~J0836-43 in comparison to the Galactic reflection nebula NGC 7023 (green) with the low resolution {\it Spitzer} spectrum of NGC 7023 scaled to match HIZOA~J0836-43 at 10\micron.}
\label{7023}
\end{center}
\end{figure}

Since HIZOA~J0836-43 appears to have dust emission characteristics similar to that of NGC 7023, a reflection nebula, we explore this similarity further and plot in Fig. \ref{7023} the low resolution (SL/LL) spectrum of NGC 7023 scaled to match HIZOA~J0836-43 at 10\micron.
As we would expect, the composition of the two regions is quite different with HIZOA~J0836-43 showing strong emission from ionic species. However, the PAH emission of NGC 7023 is remarkably similar to that of HIZOA~J0836-43, further confirming our hypothesis of PDR-dominated emission. The most noticeable difference between these two systems is that for $\lambda > 20$\micron, NGC 7023 has a steeply rising continuum indicating far more warm dust emission compared to HIZOA~J0836-43. This arises from the \HII\ region producing the reflection nebula. It is apparent that the PAH emission from HIZOA~J0836-43 closely resembles that of a reflection nebula such as NGC 7023, yet the dust grain composition and environment in HIZOA~J0836-43 is clearly different, seen most notably in the weak dust continuum of the galaxy.

The broadband photometry of HIZOA~J0836-43, with an excess of 8\micron\ and 70\micron\ emission relative to 24\micron\ and 160\micron\ (Paper I), is consistent with the strong PAH emission bands and weakly rising continuum seen in the galaxy's spectrum. The spectroscopy and photometry independently indicate this distinguishing trait of HIZOA~J0836-43; the origin of this dissimilarity compared to other local extragalactic objects poses a challenge to explain.

The extended star formation in HIZOA~J0836-43 was discussed in Paper 1 by comparing the stellar light distribution with the PAH emission surface brightness (see Fig. 1); the galaxy appears to be building its disk from the inside out. This gives rise to the increasing PAH luminosities when we compare the nuclear region spectrum with the larger 37\arcs\ extraction (Appendix A). For example,  the total PAH luminosity increases from $L_{\rm{PAH}} \sim 3.2 \times 10^{9} L_{\odot}$ in the nuclear region to $ \sim 6.5 \times 10^{9} L_{\odot}$ in the larger extraction (measured using a spline continuum).  The $L_{\rm{PAH}}/L_{\rm{TIR}}$ ratio increases from $\sim 3\%$ to $\sim 6\%$ as we include more of the disk (or from $\sim 9\%$ to $\sim 20\%$ if we use the PAHFIT values; see Appendix B). However, we note that the 37\arcs\ extraction only includes the nuclear and $inner$ disk. We measure the extent of the PAH emission (using the 11.3\micron\ PAH map; see Fig. 1) and determine an angular size of $\sim$ 65\arcs, asymmetrically distributed to the East of the nuclear region. This corresponds to a star forming disk $>$50\,kpc in diameter. The galaxy has a substantial stellar mass in place ($M_{\rm stellar} = 4.4 \times 10^{10} M_\sun$) with a prominent NIR bulge. Given the fuel available ($M_{\rm{HI}} = 7.5\times10^{10}\, M_{\sun}$) and the active star formation ($\sim 21\, M_{\odot}\, \rm{yr^{-1}}$), we are observing a galaxy in an uncommon evolutionary state, i.e. as it vigorously builds a massive stellar disk.

\section{Discussion - A Disk from the past?}

In the local universe ($z<0.1$), LIRGS are relatively rare comprising $< 5$\% of the total infrared energy density of local galaxies \citep{Soif91}. These systems are dominated by galaxy mergers and strong interactions, as evidenced by their disturbed morphologies 
\citep[see e.g.][]{Far01,Ish04}. 

Compared to the systems in the GOALS (Great All-Sky LIRG Survey) study \citep{Ar09}, HIZOA~J0836-43 has a low $L_{24 \mu {\rm m}}/L_{70 \mu {\rm m}}$ ratio and a weaker MIR continuum than the ``coldest'' LIRG in the sample, NGC 5734 (Paper I). This is an S0 galaxy ($L_{\rm{TIR}} = 1.20 \times  10^{11} L_{\odot}$) that it is interacting with the galaxy, NGC 5743, triggering the observed star formation.

LIRGS are more common at higher redshifts with $\sim$ 70\% of the infrared background at $z=1$ being produced by LIRGS \citep{LeF05}, but $\sim$ 50\% of intermediate redshift ($0.1 < z < 1$) LIRGS have disks that show little evidence of a recent merger 
\citep{Bell05,Mel08}. 
These LIRGs, which are predominantly gas-rich spiral systems, could be triggered by weak interactions with neighbours, minor mergers and internal bar instabilities, achieving heightened star formation as a result of higher gas fractions compared to local disks \citep{Mel05}. 



We speculate that HIZOA~J0836-43's current level of active star formation comes after a relatively long period of inactivity in which the galaxy would have resembled an S0/Sa type (much as Malin 1 appears today). HIZOA~J0836-43 appears to have neighbouring dwarf galaxies \citep[][Paper III; in preparation]{Clu09}, but no major merger interactions. As noted earlier, the infrared emission exhibits an asymmetry, or warp, along the eastern side in the PAH spectral map and this irregularity appears discernible in the 20-cm radio continuum and \HI\ channel maps. This feature is possibly the result of recent accretion (such as a minor merger) or tidal interaction that has exacerbated a disk instability, which may have triggered the starburst as gas flows into the dense centre (Paper I).

The study of \citet{Rob09} has shown that major interactions/mergers were an insignificant factor in stellar mass growth for $z<1$, but are, however, crucial in triggering the most intense starbursts. This emphasises the relevance of systems such as HIZOA~J0836-43, which are undergoing increased star formation without the presence of a major merger.

\begin{figure}[!thbp]
\begin{center}
\epsfxsize=8cm
\epsffile{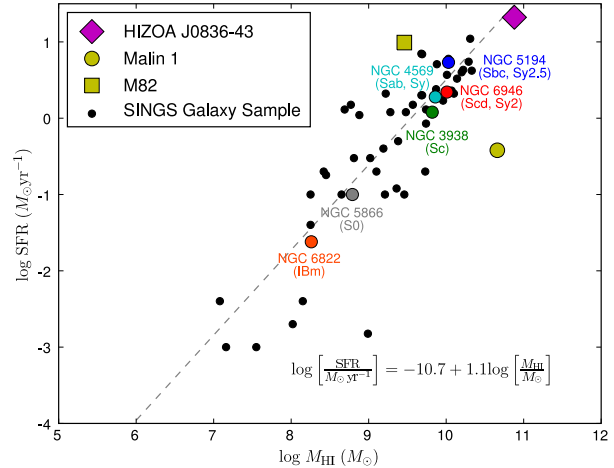}
\caption[]{SFR versus \HI\ mass for HIZOA~J0836-43 and SINGS galaxies. The star formation rates for the SINGS galaxies are calculated from their H$\alpha$ emission \citet{Ken03}. The SFR of Malin 1 is from \citet{Rah07} and its \HI\ mass from \citet{Pick97}. The M82 data is from \citet{Yun93} and \citet{San03} for the \HI\ and SFR respectively.}
\label{fig:1}
\end{center}
\end{figure}

Gas-rich galaxies are expected to be more common in the distant universe and evidence suggests that gas consumption, and not major mergers, were driving stellar mass growth \citep{Dad08, Mel08}.
Compared to local ``normal'' galaxies, HIZOA~J0836-43 appears to be a ``scaled-up'' disk galaxy in terms of its gas mass and star formation properties. This is illustrated in Figure \ref{fig:1} plotting SFR versus \HI\ mass for galaxies in SINGS, including here Malin 1, M82 and HIZOA~J0836-43. A linear least squares fit through the SINGS points shows that HIZOA~J0836-43 lies on the high-end of this relation, as opposed to Malin 1 which is quiescent by comparison and M82, the prototypical starburst. As discussed in Paper 1, the combined photometric SED (Spectral Energy Distribution) is best fit by an Sc-type galaxy template dominated by cold dust emission longwards of 60\micron, although with stronger PAH and dust emission, and does not match a starburst template.


The {\it Spitzer} IRS spectrum of HIZOA~J0836-43 shows weak silicate absorption indicating that the total (internal) dust extinction is low. The galaxy shows increased PAH emission compared to local spirals, yet preserves a relatively weak warm dust continuum emission. This suggests that the starburst geometry is extended, reducing the total dust column compared to compact or embedded star-forming regions. This agrees with the extended star formation observed in the disk (see section \ref{mir_prop}). Alternatively, if regions of star formation are gas-rich, the total dust extinction could be less as the same level of star formation is sustained for a smaller dust mass. These scenarios were similarly suggested by \citet{Far08} to explain distant ($z\sim 1.7$) ULIRG spectra that appear more similar to local starbursts than local ULIRGs.

Recent work by \citet{Dad10a} has shown that normal, near-infrared selected ($BzK$) galaxies at $z\sim1.5$ have very high gas fractions and can be viewed as scaled-up local disk galaxies hosting low efficiency star formation in extended, low excitation gas disks. Using their $L_{\rm IR}$ vs $L'_{\rm CO}$ relation, including local spirals and $BzK$ galaxies, the $L'_{\rm CO}$ lower limit for HIZOA~J0836-43 ($8.4 \times 10^8$ K\,km\,s$^{-1}$\,pc$^2$) lies on the relation exactly between these populations. 
Gas mass fractions ($M_{\rm gas}/(M_{\rm gas}+M_{\rm stellar}$) with $M_{\rm gas}$ including \HI, H$_2$ and He  are crucial for comparisons with local spirals and more distant systems, such as the $BzK$s.
Determining the gas mass fraction using our current CO data yields $\sim$64\%, comparable to the $BzK$ galaxies ($\sim$57\%) and far above the average of 20\% for local spirals (Leroy et al. 2008), despite being a lower limit.

We calculate a star formation efficiency (SFE = $L_{\rm IR}/L'_{\rm CO}$) for HIZOA~J0836-43 of 140 $L_\sun$ (K\,km\,s$^{-1}$\,pc$^2$)$^{-1}$, comparable to the average SFE for $BzK$ galaxies of $\approx$ 100 $L_\sun$ (K\,km\,s$^{-1}$\,pc$^2$)$^{-1}$ found by Aravena et al. (2010). This is below the average determined for high-redshift sub-millimeter galaxies, or SMGs, (560$\pm$210 $L_\sun$ (K\,km\,s$^{-1}$\,pc$^2$)$^{-1}$; Tacconi et al. 2006) and local ULIRGs ($\sim$ 225 $L_\sun$ (K\,km\,s$^{-1}$\,pc$^2$)$^{-1}$; Yao et al. 2003). The SFE is, however, similar to values found for local spirals (Leroy et al. 2008) and for other disk galaxies at high redshift (Tacconi et al. 2010)

Similarly, \citet{Dan09} find evidence of low-excitation molecular gas reservoirs in NIR-selected massive galaxies at $z\sim1.5$. The galaxies have star formation efficiencies and CO excitation properties similar to local spirals and the Milky Way; using a Milky Way molecular conversion factor yields molecular gas reservoirs of $\sim 10^{11} M_{\odot}$ and gas mass fractions of $\ge 0.6$. They speculate that the low-excitation properties observed may be typical in distant galaxies.

\citet{Tac10} have shown that normal, massive star forming $z\sim1.2$ and $z\sim2.3$ are gas rich with high molecular gas fractions. In terms of SFR and gas mass, they are scaled-up versions of normal $z\sim0$ gas rich disk galaxies. Compared to this sample, HIZOA~J0836-43 has a molecular gas fraction at the low end of the $z\sim1.2$ galaxies, but does not appear similar to the $z\sim2.3$ population. In addition, HIZOA~J0836-43 lies on the SFR-$M_{\rm stellar}$ correlation determined using the $z\sim1.2$ sample, but shows no agreement with the $z\sim2.3$ fit. 

This SFR-$M_{\rm stellar}$ relation was observed by Noeske et al. (2007) who found that for 2905 galaxies in the AEGIS (All-wavelength Extended Groth Strip International Survey) field there exists a ``main sequence'' where a given redshift corresponds to a limited range of SFRs and $M_{\rm stellar}$. Combining samples at different redshifts from a number of studies yields the following ``main-sequence line'' correlation between SFR and stellar mass as a function of redshift: \[ {\rm SFR} (M_{\sun}\, \rm{yr^{-1}}) = 150(M_{\rm stellar}/10^{11}M_{\sun})^{0.8}([1+z]/3.2)^{2.7} ]\] (Bouch\'{e} et al. 2010). Substituting HIZOA~J0836-43's SFR and stellar mass, corresponds to a redshift of $z\sim0.95$.

There is an emerging picture of two evolutionary tracks in observed galaxy star formation, one merger-driven, the other purely contained in star forming disks \citep{Gen10}. \citet{Dad10b} propose two modes of star formation, a long-lasting mode for disks and a more rapid mode for starbursts. They attribute this to the fraction of molecular gas in dense clouds. We find that HIZOA~J0836-43 lies on the sequence of disks (when comparing $L_{\rm IR}$ vs $M_{{\rm H_2}}$ and $L_{\rm IR}$/$M{_{\rm H_2}}$ vs $L_{\rm IR}$), scaled-up from spirals and in the locus occupied by $BzK$ galaxies.

A study of 181 LIRGs at intermediate redshift ($z\sim 0.7$) by \citet{Gio10} find the median stellar mass of the sample to be $M_{\rm stellar} \sim 6.3\times10^{10}\, M_{\sun}$ and a median SFR of $23\, M_{\sun}\, \rm{yr^{-1}}$. HIZOA~J0836-43's stellar mass ($M_{\rm stellar} \sim 4.4\times10^{10}\, M_{\sun}$) and SFR ($\sim 21\, M_{\sun}\, \rm{yr^{-1}}$ from the FIR photometry) are comparable to these distant systems. This combined with the gas fractions and star formation efficiencies previously discussed, suggest a tantalising link to a distant epoch of star formation. 

HIZOA~J0836-43 appears unusual in the local universe as an \HI-massive LIRG. However, many galaxies have large reservoirs of \HI\ that are not currently being converted into stars, and will indeed lie dormant unless the density threshold of neutral gas is exceeded, enabling molecular clouds to form and subsequently collapse.  A recent event has likely triggered the current starburst, whereas similarly extreme \HI-massive disks observed in the local universe remain quiescent. This could be due to a combination of their rarity and the `catch 22' situation that the low density environment that has preserved them \citep{Don06} has also left them bereft of suitable triggers. We could speculate as to how HIZOA~J0836-43 formed its huge gas disk -- the possibility that it is the product of a gas-rich merger of two galaxies, or due to accretion of cold gas from the intergalactic medium -- and the effect of nurture due to its environment (explored further in Paper III). But, irrespective of its origin, we are observing inside-out stellar disk-building in a gas-dominated galaxy, with minimal contamination from an AGN. If this galaxy mimics the pristine, early evolution of a gas disk, so critical to our understanding of the complex systems that subsequently form, HIZOA~J0836-43 could provide a singular opportunity for studying a star formation mechanism (at $z\sim0.036$), which is pervasive in the $z\sim 1$ universe.

\vfill

\section{Summary}

In this paper we have presented the results of our {\it Spitzer} infrared and molecular gas observations of HIZOA J08360-43. We summarise here the main findings:

\begin{itemize}

\item{The galaxy spectrum appears dominated by strong, extended PAH emission with $L_{\rm{PAH}} \sim 3.2 \times 10^{9} L_{\odot}$ in the nuclear region increasing to $ \sim 6.5 \times 10^{9} L_{\odot}$ for a region including the inner disk (measured using a spline continuum).  This corresponds to $L_{\rm{PAH}}/L_{\rm{TIR}}$ ratio of $\sim 3\%$ (nuclear) to $\sim 6\%$ (nuclear+inner disk).}

\item{In strong contrast, however, the dust 24\micron\ continuum appears intrinsically weak compared to other starburst systems and (U)LIRGs, resembling that of an ``average'' disk galaxy. Using the synthetic 24\micron\ flux, integrated from the nuclear region spectrum, corresponds to $L_{\rm PAH}$/$L_{24}$ of $\sim$ 1.3. This paucity of emission from warm dust (indicating a lack of emission from very small grains) is seen independently in the galaxy's photometry.}

\item{Compared to LIRGs with similar infrared luminosity ($L_{\rm TIR}$=10$^{11}L_{\sun}$), the PAHs in HIZOA~J0836-43 are on average $\sim$2 times stronger, whereas the warm dust continuum ($\lambda > 20$\micron) is best fit by a star forming galaxy with $L_{\rm TIR}$=10$^{10}L_{\sun}$.}

\item{The combination of strong PAH emission and relatively weak continuum places HIZOA~J0836-43 in the extreme PDR-dominated region of the MIR diagnostic diagram of Peeters et al. (2004), similar to the reflection nebula NGC 7023. We speculate that the absence of a substantial warm dust continuum may be due to the dust geometry in a gas-rich disk, as the star formation appears strongly extended across a region of $>$50\, kpc and not severely embedded.}

\item{The $[$Ne\,{\sc iii}$]$\,15.56\micron/$[$Ne\,{\sc ii}$]$\,12.81\micron\ ratio indicates a typical radiation field strength compared to other starbursting systems and the electron density of $n_{e} \sim 300\, \rm{cm}^{-3}$ (from the $[$S\,{\sc iii}$]\, 18.71$\micron/$[$S\,{\sc iii}$]\, 33.48$\micron\ ratio) is also normal for disk systems. The $[$Si\,{\sc ii}$]$\,34.82\micron/$[$S\,{\sc iii}$]$\,33.48\micron\ ratio suggests strong PDR emission compared to emission from \HII\ regions.}

\item{We determine a lower limit for the molecular gas content of the galaxy ($\sim$4$\times 10^{9} M_\sun$); the corresponding gas mass fraction is $\sim$64\%, comparable to the $BzK$ galaxies ($\sim$57\%) and far above the average of 20\% for local spirals. However, the star formation efficiency (SFE = $L_{\rm IR}/L'_{\rm CO}$) for HIZOA~J0836-43 of 140 $L\sun$ (K\,km\,s$^{-1}$\,pc$^2$)$^{-1}$ is similar to that of local spirals, $BzK$ galaxies and other disk galaxies at high redshift.}

\item{HIZOA~J0836-43 appears scaled-up compared to local spirals in terms of its SFR, stellar mass and gas content. The combination of SFR and stellar mass suggests that it would be considered typical at a redshift of $z\sim1$ and its stage of active building, given its gas supply, appears analogous to these distant systems.\\}

\end{itemize}

\vspace{1cm}

\acknowledgements

We thank Danny Dale and SINGS, Lee Armus and GOALS for data access. We are grateful to Steve Lord, Ray Norris, Joe Mazzarella and Justin Howell for valuable discussion and input. We thank the ATNF for hosting us (in Epping and Narribri), and in particular Balt Indermuehle for his input regarding our Mopra observations. Support for this work was provided by NASA through an award issued by JPL/Caltech. M.E.C, R.K-K and P.A.W thank the NRF (South Africa) for financial support.

\appendix

\section*{Appendix A: Emission Lines in $37^{\prime \prime}$ SL Extraction}\label{SL}

\begin{figure}[!tbh]
\begin{center}
\includegraphics[width=8cm]{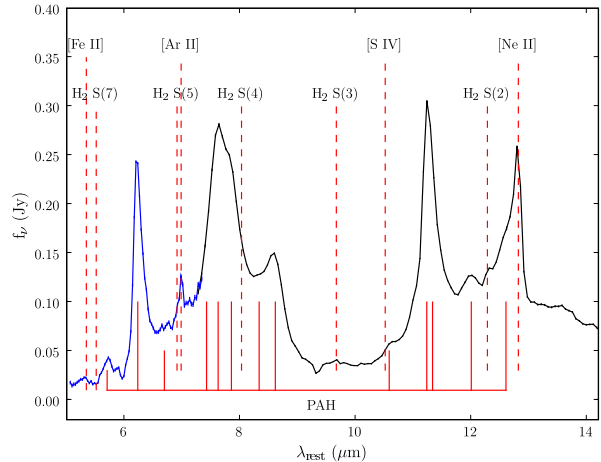}
\caption{IRS SL spectrum of HIZOA~J0836-43; 37\arcs\ aperture. The blue line represents the SL2 module and the black line the SL1 module.}
\label{fig:sl}
\end{center}
\end{figure}

The {\it Spitzer} SL mapping coverage enables us to make a large ($37^{\prime \prime}$ aperture) extraction of HIZOA~J0836-43 which includes the nuclear region and the inner part of the MIR disk. Given the short wavelength coverage ($5-14$\micron) we measure predominantly PAH emission, as shown in Figure \ref{fig:sl}. Table \ref{tab:SL} contains the measurements made using a spline continuum.

Summing the individual luminosities for each feature corresponds to a total PAH luminosity of $L_{\rm{PAH}} \sim 6.5 \times 10^{9} L_{\odot}$, compared to that obtained for the 9.25\arcs\ nuclear extraction in Section \ref{pah} ($L_{\rm{PAH}} = 3.2 \times 10^{9} L_{\odot}$). The $L_{\rm{PAH}}/L_{\rm{TIR}}$ ratio therefore increases (compared to the nuclear extraction) to 0.06 or 6\% from $\sim 2.7\%$.

\begin{table}[!ht]
{\scriptsize
\begin{minipage}{16cm}
\caption{Emission line measurements - $37^{\prime \prime}$ Aperture (SL).  \label{tab:SL}}
\begin{center}
\begin{tabular}{c r r r r r}
\hline
\hline
\\[0.5pt]
  Line     &  $\lambda_{rest} $   & FWHM  &  EQW   & Flux $^{a}$ &  $\nu L_{\nu}$ $^{a}$  \\

 &    $\mu \rm{m}$       &  $10^{-3}\mu \rm{m}$    &    $   \mu \rm{m}$    &   $10^{-17} \ \rm{W} \ \rm{m}^{-2}$ &  $10^{7} L_{\odot}$  \\
\hline
\\
PAH & 6.23    &    &     0.996   &    295.1 $\pm 6.9$  &      201.3 $\pm 4.7$
\\
$[$Ar\,{\sc ii}$]$  $^{b}$ &   6.98   &     88.2   &  0.043    &   21.17 $\pm 1.57$   &      14.4 $\pm 1.1$
\\
PAH  &   7.67   &     &   0.744    &   423.2 $\pm 9.5$  &     288.6 $\pm 6.4$
\\
PAH   &   8.61   &    &   0.106   &   59.87 $\pm 1.27$   &        40.8 $\pm 0.9$
\\
PAH   &   11.26   &     &  0.735   &   141.2  $\pm 2.2$  &    96.3 $\pm 1.5$
\\
$\rm{H}_{2}$ {\it S}(2)  &  12.28  & 200.0   &   0.157   &  4.54  $\pm 0.91$   &  3.1 $\pm 0.6$
\\
PAH   &   12.60   &      &   0.361    &   33.10 $\pm 1.14$  &    22.6 $\pm 0.8$
\\
$[$Ne\,{\sc ii}$]$   &   12.81   &   155.0  &  0.281   &   40.47 $\pm 0.71$ & 27.6 $\pm 0.5$
\\
\\
\hline

\\[0.5pt]

\multicolumn{6}{l}{$^{a}$ Error corresponds to line flux relative to baseline RMS.}\\
\multicolumn{6}{l}{$^{b}$ [Ar II] is blended with $\rm{H}_{2}$ {\it S}(2).}\\
\end{tabular}
\end{center}
\end{minipage}
}
\end{table}

\section*{Appendix B: PAHFIT Measurements}\label{pahfit}

An alternative method for measuring PAH emission is to robustly disentangle the emission lines, dust features (e.g. PAH complexes), silicate absorption and dust continuum components, as described in \citet{Smi07}, using PAHFIT an IDL package developed to decompose IRS low-resolution spectra. We summarise in Table \ref{tab:pahf} the PAH equivalent widths and luminosities obtained using PAHFIT on the nuclear region extraction of HIZOA~J0836-43; these can be used to compare to PAH studies that utilise this method \citep[for example][]{Smi07}.

\begin{table}[!htp]
{\scriptsize
\begin{minipage}{16cm}
\caption{PAHFIT Emission Line Equivalent Widths and Luminosities (nuclear region). \label{tab:pahf}}
\begin{center}
\begin{tabular}{c c r r }
\hline
\hline
\\[0.5pt]
   PAH $\lambda_{rest} $       &  EQW  &  Flux   & Luminosity\\

        $\mu \rm{m}$           &    $  \mu \rm{m}$   &   $10^{-17} \ \rm{W} \ \rm{m}^{-2}$       &  $(10^{7} L_{\odot})$ \\
\hline
\\
5.27    &   0.186    &   12.7   &  8.7
\\
5.70    &   0.211    &   16.9   &  11.5
\\
6.22    &   2.030    &   182.0   &  124.2
\\
6.69    &   0.489    &   45.9   &  31.3
\\
7.42   &   1.420    &   132.0   &  90.1
\\
7.60   &   2.760    &   255.0   &  174.0
\\
7.85   &   2.630    &   236.0   &  161.0
\\
8.33    &   0.656    &   53.5   &  36.5
\\
8.61    &   1.610    &   119.0   &  81.2
\\
10.68   &   0.179    &   8.2   &  5.59
\\
11.23    &   1.020    &   48.0   &  32.8
\\
11.33   &   2.230    &   106.0   &  72.3
\\
11.99    &   0.975    &   48.0   &  32.8
\\
12.62   &   1.700    &   81.7   &  55.7
\\
12.69   &   0.097    &   4.66   &  3.2
\\
13.48   &   0.527    &   23.7   &  16.2
\\
14.19    &   0.124    &   5.31   &  3.6
\\
16.45   &   0.224    &   8.03   &  5.5
\\
17.04   &   0.761    &   26.1  &  17.8
\\
17.38   &   0.151    &   5.02   &  3.4
\\

\hline
\\[0.5pt]
\\[0.5pt]
\end{tabular}
\end{center}
\end{minipage}
}
\end{table}

Summing the individual luminosities for each feature corresponds to a total PAH luminosity of $L_{\rm{PAH}} \sim 9.7 \times 10^{9} L_{\odot}$, approximately three times the luminosity obtained in Section \ref{pah} ($L_{\rm{PAH}} = 3.2 \times 10^{9} L_{\odot}$) using the spline continuum measurements. This corresponds to a $L_{\rm{PAH}}/L_{\rm{TIR}}$ ratio of 0.09 or 9\%. We compare this result to the study of \citet{Smi07} who used the PAH decomposition method to investigate PAH emission in 59 starforming galaxies in SINGS, determining that the majority of galaxies have $L_{\rm{PAH}}/L_{\rm{TIR}}$ close to 0.1, with lower and upper limits of $\sim0.02$ and  0.11, respectively. However, given the extended star formation in HIZOA~J0836-43's ratio of 0.06 is on the low side compared to the majority of their galaxies, especially since it is a LIRG with a relatively high SFR.

Table \ref{tab:pahcomp} compares the luminosities of the strongest PAH complexes (6.2, 7.7 and 11.3\micron), measured for the nuclear region, to the luminosities obtained using the spline continuum (from Section \ref{pah}). The $L$(decomp)/$L$(spline) ratios are similar to what has been found for star forming galaxies in SINGS \citep{Smi07} as shown in Table \ref{tab:pahcomp}.

\begin{table}[!hp]
{\scriptsize
\begin{minipage}{16cm}
\caption{Comparison between PAH feature luminosities - PAHFIT versus Spline Continuum \label{tab:pahcomp}}
\begin{center}
\begin{tabular}{c r c c }
\hline
\hline
\\[0.5pt]
   PAH        &  $L_{\rm pahfit}$   &  $\frac{L_{\rm pahfit}}{L_{\rm spline}}$ &  $\frac{L_{\rm pahfit}}{L_{\rm spline}}$ $^a$\\

                              &  $(10^{8} L_{\odot})$   &   &  \\
\hline
\\
6.2\micron    &   12.42    &  1.86  & 1.70
\\
7.7\micron\ $^b$   &  42.5      &  2.83 & 3.53
\\
11.3\micron\ $^b$   &   10.51  &  2.58  & 1.86
\\
\hline
\\[0.5pt]
\multicolumn{4}{l}{$^{a}$ \citet{Smi07}} \\
\multicolumn{4}{l}{$^{b}$ PAH complex} 
\end{tabular}
\end{center}
\end{minipage}
}
\end{table}

Using PAHFIT on the large extraction (37\arcs), shown in Figure \ref{fig:sl}, allows us to measure the PAH emission contained in the nuclear redgion and the inner MIR disk. The results are summarised in Table \ref{tab:pahf_big}.

\begin{table}[!htp]
{\scriptsize
\begin{minipage}{16cm}
\caption{PAHFIT Emission Line Equivalent Widths and Luminosities (37\arcs\ Extraction). \label{tab:pahf_big}}
\begin{center}
\begin{tabular}{c c r r }
\hline
\hline
\\[0.5pt]
   PAH $\lambda_{rest} $       &  EQW  &  Flux   & Luminosity\\

        $\mu \rm{m}$           &    $  \mu \rm{m}$   &   $10^{-17} \ \rm{W} \ \rm{m}^{-2}$       &  $(10^{7} L_{\odot})$ \\
\hline
\\
5.27    &   4.060    &   58.9   &  40.2
\\
5.70    &   3.070    &   73.6   &  50.2
\\
6.22    &   13.50    &   517.0   &  352.6
\\
6.69    &   4.020    &   217.0   &  148.0
\\
7.42   &   5.970    &   470.0   &  320.5
\\
7.60   &   5.330    &   442.0   &  301.4
\\
7.85   &   4.910    &   448.0   &  305.5
\\
8.33    &   1.290    &   132.0   &  90.0
\\
8.61    &   2.050    &   219.0   &  149.4
\\
10.68   &   0.085    &   11.1   &  7.57
\\
11.23    &   0.660    &   93.9   &  64.0
\\
11.33   &   1.390    &   201.0   &  137.1
\\
11.99    &   0.583    &   89.5   &  61.0
\\
12.62   &   0.936    &   147.0   &  100.3
\\
12.69   &   0.104    &   16.3   &  11.1
\\
13.48   &   0.205    &   31.5   &  21.5
\\

\hline
\\[0.5pt]
\\[0.5pt]
\end{tabular}
\end{center}
\end{minipage}
}
\end{table}

The total PAH luminosity in the 37\arcs\ extraction is $L_{\rm{PAH}} \sim 2.2 \times 10^{10} L_{\odot}$, corresponding to $L_{\rm{PAH}}/L_{\rm{TIR}}$ ratio of 0.2 or 20\%. The $L_{\rm{PAH}}/\nu L_{\nu}(24\mu m)$ ratio is $\sim 2$; this corresponds to twice as much power coming from PAH features compared to warm dust emission. Using the PAH luminosities for the two extractions listed in Table \ref{tab:pahf} and \ref{tab:pahf_big}, we can calculate $L_{\rm{PAH}}/\nu L_{\nu}(24\mu m)$ ratios of the 6.2, 7.7 and 11.3\micron\ features. We list these in Table \ref{tab:pahfin}, as well as values from the study of \citet{Smi07}.

Despite the luminosity ratio's in \citet{Smi07} showing a large scatter, the large extraction of HIZOA~J0836-43 exceeds the values obtained for the SINGS sample. For example, the $L_{\rm{6.2\mu m}}/\nu L_{\nu}(24\mu m)$ and $L_{\rm{11.3\mu m}}/\nu L_{\nu}(24\mu m)$ exceed the highest ratios seen in star forming galaxies in SINGS \citep{Smi07}. Given that we are not extracting over the entire star forming disk, we expect this ratio to be even larger.

\begin{table}[!htbp]
{\scriptsize
\begin{minipage}{16cm}
\caption{Ratios of PAH luminosity to 24\micron\ luminosity \label{tab:pahfin}}
\begin{center}
\begin{tabular}{c r r}
\hline
\hline
\\[0.5pt]
   Ratio       &  $L_{\rm{PAH}}/\nu L_{\nu}(24\mu m)$   &    $L_{\rm{PAH}}/\nu L_{\nu}(24\mu m) ^{a}$   \\

\hline
\\
$L_{\rm{6.2\mu m}}/\nu L_{\nu}(24\mu m)$ & $\sim 0.11$ $^b$   & $\sim 0.1$
\\
$L_{\rm{7.7\mu m}}/\nu L_{\nu}(24\mu m)$ &  $\sim 0.46$ $^b$  & $ \sim 0.4$
\\
$L_{\rm{11.3\mu m}}/\nu L_{\nu}(24\mu m)$   &   $\sim 0.10 $ $^b$ &  $\sim 0.1$
\\
$L_{\rm{6.2\mu m}}/\nu L_{\nu}(24\mu m)$ & $\sim 0.32$  $^c$  & $\sim 0.1$
\\
$L_{\rm{7.7\mu m}}/\nu L_{\nu}(24\mu m)$ &  $\sim 0.84$  $^c$ & $ \sim 0.4$
\\
$L_{\rm{11.3\mu m}}/\nu L_{\nu}(24\mu m)$   &   $\sim 0.18 $ $^c$ &  $\sim 0.1$
\\

\hline
\\[0.5pt]
\multicolumn{3}{l}{$^{a}$ \citet{Smi07}} \\
\multicolumn{3}{l}{$^{b}$ Nuclear region } \\
\multicolumn{3}{l}{$^{c}$ 37\arcs\ extraction} \\

\\[0.5pt]
\end{tabular}
\end{center}
\end{minipage}
}
\end{table}

\end{document}